\documentclass[12pt,letterpaper]{article}
\usepackage{amsmath,amssymb,array,calc,rotating,epsfig,psfrag,amscd, cite}

\makeatletter
\renewcommand\section{\@startsection {section}{1}{\z@}%
                                   {-3.5ex \@plus -1ex \@minus -.2ex}
                                   {2.3ex \@plus.2ex}%
                                   {\normalfont\large\bfseries}}
\renewcommand\subsection{\@startsection{subsection}{2}{\z@}%
                                     {-3.25ex\@plus -1ex \@minus -.2ex}%
                                     {1.5ex \@plus .2ex}%
                                     {\normalfont\bfseries}}
\makeatother

\let\non\nonumber

\let\s=\sigma

\let\S=\Sigma

\newcommand{\bea}{\begin{eqnarray}}
\newcommand{\eea}{\end{eqnarray}}
\newcommand{\be}{\begin{equation}}
\newcommand{\ee}{\end{equation}}

\newcommand{\p}{\partial}

\newcommand{\C}[1]{$(\ref{#1})$}

\def\IZ{\relax\ifmmode\mathchoice
{\hbox{\cmss Z\kern-.4em Z}}{\hbox{\cmss Z\kern-.4em Z}}
{\lower.9pt\hbox{\cmsss Z\kern-.4em Z}} {\lower1.2pt\hbox{\cmsss
Z\kern-.4em Z}}\else{\cmss Z\kern-.4em Z}\fi}
\def\IR{\relax{\rm I\kern-.18em R}}

\def\one{{\hbox{ 1\kern-.8mm l}}}

\newlength{\bredde}
\def\slash#1{\settowidth{\bredde}{$#1$}\ifmmode\,\raisebox{.15ex}{/}
\hspace*{-\bredde} #1\else$\,\raisebox{.15ex}{/}\hspace*{-\bredde}
#1$\fi}

\newsavebox{\zzzbar}
\sbox{\zzzbar}
  {\setlength{\unitlength}{0.9em}
  \begin{picture}(0.6,0.7)
  \thinlines
  \put(0,0){\line(1,0){0.6}}
  \put(0,0.75){\line(1,0){0.575}}
  \multiput(0,0)(0.0125,0.025){30}{\rule{0.3pt}{0.3pt}}
  \multiput(0.2,0)(0.0125,0.025){30}{\rule{0.3pt}{0.3pt}}
  \put(0,0.75){\line(0,-1){0.15}}
  \put(0.015,0.75){\line(0,-1){0.1}}
  \put(0.03,0.75){\line(0,-1){0.075}}
  \put(0.045,0.75){\line(0,-1){0.05}}
  \put(0.05,0.75){\line(0,-1){0.025}}
  \put(0.6,0){\line(0,1){0.15}}
  \put(0.585,0){\line(0,1){0.1}}
  \put(0.57,0){\line(0,1){0.075}}
  \put(0.555,0){\line(0,1){0.05}}
  \put(0.55,0){\line(0,1){0.025}}
  \end{picture}}

\newcommand{\ena}{\end{eqnarray}}
\newcommand{\beqa}{\begin{eqnarray}}
\newcommand{\eeqa}{\end{eqnarray}}

\newcommand{\g}{\gamma}

\def\g{\gamma}

\def\s{\sigma}

\def\S{\Sigma}

\begin{document}
\begin{titlepage}

\begin{center}



\vskip 2 cm
{\Large \bf Constraining gravitational interactions in the M theory effective action}\\
\vskip 1.25 cm { Anirban Basu\footnote{email address:
    anirbanbasu@hri.res.in} } \\
{\vskip 0.5cm Harish--Chandra Research Institute, Chhatnag Road, Jhusi,\\
Allahabad 211019, India\\}

\end{center}

\vskip 2 cm

\begin{abstract}
\baselineskip=18pt

We consider purely gravitational interactions of the type $D^{6n} \mathcal{R}^4$ in the effective action of M theory in 11 dimensional flat spacetime. The duality between M theory on $S^1$ and type IIA string theory relates them to the type IIA interactions of the form $e^{2n\phi_A} D^{6n} \mathcal{R}^4$ where $\phi_A$ is the type IIA dilaton. The coefficients of the M theory interactions are determined by the strongly coupled type IIA theory. Given the nature of the dilaton dependence, it is plausible that for low values of $n$, the coefficient has a similar structure as the genus $(n+1)$ string amplitude of the type IIA $D^{6n} \mathcal{R}^4$ interaction, namely the transcendental nature. 
Assuming this, and focussing on the even--even spin structure part of the type IIA string amplitude, this coefficient is given by the type IIB genus $(n+1)$ amplitude, which we constrain using supersymmetry, S--duality and maximal supergravity. The source terms of the Poisson equations satisfied by the S--duality invariant IIB couplings play a central role in the analysis. This procedure yields partial contributions to several multi--loop type IIB string amplitudes, from which we extract the transcendental nature of the corresponding M theory couplings. For $n \leq 2$, all possible source terms involve only BPS couplings.  While the $\mathcal{R}^4$ and $D^6\mathcal{R}^4$ M theory couplings agree with known results, the coefficient of the $D^{12} \mathcal{R}^4$ interaction takes the form $\zeta (2)^3 \Big( \Omega_1+\Omega_2 \zeta (3) \Big)$. We also analyze the $D^{18} \mathcal{R}^4$ and $D^{24} \mathcal{R}^4$ interactions, and show that their coefficients have at least the terms $\zeta (2)^4 \Big( \tilde\Omega_1 +\tilde\Omega_2\zeta (3) +\tilde\Omega_3\zeta (5)\Big)$ and $\zeta (2)^5 \Big( \underline\Omega_1 +\underline\Omega_2\zeta (3) +\underline\Omega_3\zeta (5) + \underline\Omega_4\zeta (3)^2 + \underline\Omega_5\zeta (7) +\underline\Omega_6\zeta (3) \zeta (5)  +\underline\Omega_7 \zeta (3)^3\Big)$ respectively. The various undetermined constants have vanishing transcendentality.

\end{abstract}

\end{titlepage}


\section{Introduction}

A detailed knowledge of M theory is important for our understanding of quantum gravity. 
Apart from this, it plays a central role in
providing a non--perturbative definition of string theory, and also for analyzing the consequences of the various string dualities. While little is known about the detailed dynamics of M theory, one can try to construct the effective action of M theory in various backgrounds to learn about the structure of the theory. The low energy effective action is the action of the massless modes of the theory, and is an appropriate description at long distances. The various local as well as non--local interactions in the effective action encode important information about the underlying structure and symmetries of M theory. 

While constructing the effective action of M theory in arbitrary backgrounds is difficult, some statements can be made for backgrounds which preserve enough symmetries so that it is amenable to a systematic analysis. We shall consider certain local terms in the effective action of M theory in 11 flat spacetime dimensions. These local interactions are a subset of the purely gravitational interactions of the form $D^{2k} \mathcal{R}^4$ for $k \geq 0$, and yield corrections to the Einstein--Hilbert action. At the eight derivative level, the $\mathcal{R}^4$ and the $C_3 \wedge X_8$ interactions have been analyzed in~\cite{Vafa:1995fj,Duff:1995wd,Green:1997di}. Various properties of the higher derivative corrections to the Einstein--Hilbert action have been analyzed in~\cite{Russo:1997mk,Deser:1998jz,Howe:2003cy,Cederwall:2004cg,Peeters:2005tb,Rajaraman:2005ag,Damour:2005zb,Green:2005ba,Green:2006gt,Hyakutake:2007vc}.

Given the extended supersymmetry M theory enjoys, it is expected that the various local purely gravitational interactions should be a part of the same supermultiplet at a given order in the derivative expansion for $k \leq 3$ as these interactions are BPS, and hence should have their coefficients related by supersymmetry. Thus we expect that the $\mathcal{R}^{4+k}$ interaction should be in the same supermultiplet as the $D^{2k} \mathcal{R}^4$ interaction for $k\leq 3$. For higher values of $k$, the various interactions should form a basis of distinct non--BPS supermultiplets. Hence though our analysis below will be for the $D^{2k} \mathcal{R}^4$ interaction, the same analysis follows for the various other interactions in the same supermultiplet. For the various interactions that are not in the same supermultiplet as the $D^{2k} \mathcal{R}^4$ interaction, the primary logic of the analysis is the same, though we have not considered such interactions in any detail.

In this work, we shall consider only the $D^{6n} \mathcal{R}^4$ interactions in the M theory effective action. This is because when compactified on $S^1$, they produce interactions of the form $e^{2n\phi_A} D^{6n} \mathcal{R}^4$ in the type IIA effective action, where $\phi_A$ is the type IIA dilaton. While the coefficients of these interactions in M theory are determined by the coefficients of the $e^{2n\phi_A}$ terms in the strong coupling limit of the corresponding type IIA amplitudes, given the dilaton dependence, it is plausible that for low values of $n$ (even though the type IIA interactions for $n \geq 2$ are non--BPS and are expected to receive an infinite number of perturbative contributions) the coefficients at strong coupling have a structure similar to the genus $(n+1)$ type IIA string amplitudes, namely the transcendental nature. Proceeding with this assumption, we want to analyze the transcendental nature of these perturbative string amplitudes for low values of $n$.

While this is particularly difficult to determine in the type IIA theory, we shall focus on the $t_8 t_8 R^4$ part of the $\mathcal{R}^4$ interaction, and not consider the $\epsilon_{10} \epsilon_{10} R^4$ part. Given the equality of the perturbative type IIA and type IIB string amplitudes to all genera for the $t_8 t_8 R^4$ interaction,  we focus on the specific part of the $D^{6n} \mathcal{R}^4$ interaction with this spacetime structure for the $\mathcal{R}^4$ term. Thus it is enough for our purposes to calculate the coefficients in the type IIB theory, which we constrain using S--duality and supersymmetry. 

To begin with, we briefly review the 4 graviton tree level amplitude in the type II theory, and obtain the tree level interactions at the first few orders in the momentum expansion for these spacetime interactions in the type II theory, which are relevant for our purposes. The $\mathcal{R}^4$ and $D^6 \mathcal{R}^4$ interactions in M theory are also discussed. 

In the next section, the primary method we use to determine the various genus amplitudes is discussed, following~\cite{Green:1998by,Basu:2008cf}. The constraints imposed by supersymmetry and S--duality of type IIB string theory provide strong constraints on the various genus amplitudes, which are otherwise difficult to calculate directly. The procedure to do the analysis involves exploiting the invariance of the action under supersymmetry transformations order by order in the $l_s$ expansion, which we explain along with the constraints imposed on the $\mathcal{R}^4$ and $D^6 \mathcal{R}^4$ interactions as examples. 
The coefficients of the various gravitational interactions in the type IIB effective action are given by sums of $SL(2,\mathbb{Z})$ invariant modular forms, each of which satisfies Poisson equation on the fundamental domain of moduli space. The source terms in the Poisson equations play a central role in our analysis, and determine various high genus amplitudes in the effective action. The analysis leads to the known $\mathcal{R}^4$ and $D^6 \mathcal{R}^4$ couplings in M theory. 

Generalizing the analysis to higher orders in the derivative expansion, we next consider the $D^{12} \mathcal{R}^4$ interaction, and obtain the output due to imposing the constraints. Here new issues arise due to the non--locality of the type IIB effective action which lead to contributions from the source terms which are logarithmic in the complex coupling. Our analysis of the $D^{12} \mathcal{R}^4$ interaction involves the knowledge of various BPS interactions at lower orders in the derivative expansion, which are both local as well as non--local in the external momenta. This allows up to calculate the analytic part of the genus 3 amplitude which is proportional to
\be \label{fin12} \zeta (2)^3 \Big( \Omega_1 +\Omega_2 \zeta (3)\Big).\ee
In \C{fin12} $\Omega_i$ ($i=1,2$) are numerical factors which we do not calculate. These factors have vanishing transcendentality\footnote{Hence they do not involve $\pi$ or various zeta functions, for example.}. Hence we determine the nature of the $D^{12} \mathcal{R}^4$ interaction in M theory up to numerical factors of vanishing transcendentality. 

We next consider the $D^{18} \mathcal{R}^4$ interaction schematically. This is the simplest case where the source terms in the Poisson equations involve the couplings of various non--BPS interactions at lowers orders in the derivative expansion. Due to lack of knowledge of these non--BPS couplings, the analysis gets considerably more complicated. 
However, based on the constraints of supersymmetry and S--duality of the type IIB theory, we are able to find at least a part of the coefficients of certain higher genus amplitudes which yield the transcendental nature of the $D^{18} \mathcal{R}^4$ term in the M theory effective action. 
Including only these contributions, we show that the coefficient of the $D^{18} \mathcal{R}^4$ interaction has at least the terms
\be \label{fin18}\zeta (2)^4 \Big( \tilde\Omega_1 + \tilde\Omega_2\zeta (3) +\tilde\Omega_3 \zeta (5)\Big),\ee
where the constants $\tilde\Omega_i$ have vanishing transcendentality, which we do not determine.  

Similarly, we argue that the coefficient of the $D^{24} \mathcal{R}^4$ interaction has at least the terms
\be\label{fin24}\zeta (2)^5 \Big( \underline\Omega_1 +\underline\Omega_2\zeta (3) +\underline\Omega_3\zeta (5) + \underline\Omega_4\zeta (3)^2 + \underline\Omega_5\zeta (7) +\underline\Omega_6\zeta (3) \zeta (5) +\underline\Omega_7 \zeta (3)^3\Big)\ee
where the undetermined constants $\underline\Omega_i$ have vanishing transcendentality. In \C{fin18} and \C{fin24}, we have kept the terms in an ascending order of transcendentality of the various coefficients.  

Our analysis generalizes to higher $n$ and yields partial contributions to various multi--loop type IIB amplitudes easily. However, because the source terms involve more and more non--BPS couplings as $n$ increases, the number of terms we cannot determine for the genus $n+1$ amplitude steadily increases.  Our work emphasizes the important role supersymmetry and U--dualities play in constraining the effective action in general.

\section{A class of local gravitational interactions in the M theory effective action} 

Let us analyze the $D^{2k} \mathcal{R}^4$ interactions in the M theory effective action for $k \geq 0$. Denoting the 11 dimensional Planck length by $l_{11}$, the relevant term in the effective action is given by
\be \label{actM}
S = l_{11}^{2k-3} \int d^{11} x \sqrt{-G} D^{2k} \mathcal{R}^4 ,\ee
where $G_{MN}$ is the M theory metric, and we have neglected overall numerical factors\footnote{In our entire analysis, we shall neglect an overall factor of $(4\pi)^{-8}$ that arises in \C{actM}. }. Compactifying the theory on a circle of radius $R_{11}$, and using the relations~\cite{Witten:1995ex}
\be \label{duality}l_{11} = e^{\phi_A/3} l_s, \quad R_{11}^3 = e^{2\phi_A},\ee 
where $\phi_A$ is the type IIA dilaton and $l_s$ is the string length, we see that \C{actM} yields the term in the 10 dimensional type IIA effective action given by
\be \label{actIIA}
S = 2\pi l_s^{2k-2} \int d^{10}x \sqrt{-g} e^{2k\phi_A/3} D^{2k} \mathcal{R}^4,\ee
where $g_{\mu\nu}$ is the type IIA metric in the string frame, and we have used the length element
\be \label{metric}ds^2 = G_{MN} dx^M dx^N= g_{\mu\nu} dx^\mu dx^\nu + R_{11}^2 (dx^{11} - C_\mu dx^\mu)^2.\ee
If $k=3n$, the dilaton dependence of the interaction in \C{actIIA} is $e^{2n\phi_A}$. Thus given \C{duality}, we see that the coefficient of the interaction in \C{actM} is given by the coefficient of the $e^{2n\phi_A}$ term in the type IIA theory at strong coupling. For $n=0,1$, these type IIA couplings are BPS and receive only a finite number of perturbative contributions, and hence the strong coupling limit simply picks out the genus 1 and 2 coefficients respectively\footnote{The $D^4\mathcal{R}^4$ interaction vanishes in the M theory effective action.}. The interactions for $n \geq 2$ are non--BPS and are expected to receive an infinite number of perturbative contributions, and the strong coupling limit is expected to be significantly different from the weak coupling expansion. However, for small enough of values of $n$, it is plausible that some features of the structure of the strong coupling expansion are captured by the weak coupling expansion, namely the coefficient of the genus $(n+1)$ string amplitude. By this we specifically mean the transcendentality of the coefficients that arise in the perturbative string calculation\footnote{This agrees with a particular strong coupling calculation for $n=2$, which we mention later.}. We shall assume this to be true in our discussions below. This conclusion also holds for other interactions in the same supermultiplet.

This argument does not involve the specific details of the $D^{6n} \mathcal{R}^4$ interaction, and only involves the number of derivatives. One could similarly analyze the $\mathcal{R}^{4+3n}$ interactions in the M theory effective action, along the lines of~\cite{Russo:1997mk,Damour:2005zb}.

\section{The leading terms in the M theory effective action}

Based on the discussion above, in order to analyze the nature of the coefficient of the $D^{6n} \mathcal{R}^4$ interaction in \C{actM}, we need to know the coefficient of the genus $(n+1)$ amplitude for the $D^{6n} \mathcal{R}^4$ interaction in the type IIA theory. While this is rather difficult to calculate, we shall see below that restricting ourselves to a certain spacetime structure for the $\mathcal{R}^4$ interaction, the use of the constraints imposed by supersymmetry and S--duality in type IIB string theory allows us to put some constraints on this amplitude, and correspondingly on the $D^{6n} \mathcal{R}^4$ interaction in M theory.

Including the various contributions from the terms that we have calculated for small values of $n$, our results suggest that the coefficient of the $D^{6n} \mathcal{R}^4$ interaction in M theory is given by sums of powers of various zeta functions at least for small values of $n$. Note that by zeta functions, we mean not just Riemann zeta functions, but also their generalization--the Multiple Zeta Values (MZV)\footnote{We shall exhibit an interaction in the M theory effective action whose coefficient can involve an MZV in section 4.7.}. Calculating these coefficients is not possible for generic $n$ with the our current understanding of string amplitudes or regularized maximal supergravity. This is because the various methods to do the calculation are far too intricate to yield the answer. 

So far as loop amplitudes in maximal supergravity are concerned, the techniques used in~\cite{Green:1997as,Green:1999pu,Green:2005ba,Green:2008bf} get considerably more difficult as $n$ increases, as one needs to go to higher and higher loops in 11 dimensional supergravity compactified on $S^1$ and regularize the ultraviolet divergences of the supergravity theory. Also the non--BPS interactions are expected to receive contributions from all loops in regularized supergravity, which make them particularly difficult to calculate. However, upto numerical factors of vanishing transcendentality, the zeta functions that arise for the various genus amplitudes  in string theory can be obtained from the supergravity analysis, which will be very helpful for our purposes. 

Alternatively, the method employed in~\cite{Green:1998by,Basu:2008cf} gets very involved as $n$ increases because of the intricate nature of supersymmetry beyond the linearized level. However, this will provide considerable insight into the nature of the higher genus string amplitudes which will be helpful for us, even though the absence of an off--shell formulation makes the constraints due to supersymmetry difficult to analyze. For the cases that we shall analyze, we shall see that the techniques used in~\cite{Green:1998by,Basu:2008cf} provide non--trivial information about multi--loop string amplitudes for a given interaction, which directly determines the nature of interactions in M theory. 

It is known that the 4 graviton amplitude is the same for either of the type II string theories upto genus 4~\cite{Berkovits:2006vc}. These amplitudes the not the same beyond genus 4, because the $\epsilon_{10} \epsilon_{10} R^4$ part of the amplitude coming from the odd--odd spin structure is different. However, the $t_8 t_8 R^4$ part of the amplitude which involves the contribution from the even--even spin structure, is the same for type IIA as well as type IIB string theory at arbitrary orders in the genus expansion as well as the momentum expansion, where the momentum factors act on $R^4$. Thus we shall analyze only the coefficient of the genus $(n+1)$ amplitude for the $D^{6n} t_8 t_8 R^4$ interaction. Hence it is enough for our purposes to calculate the coefficient in type IIB string theory. The reason we want to do the calculation in the type IIB theory is because we want to impose the constraints implied by S--duality which allow us to obtain explicit expressions for these amplitudes at high genera, along the lines of~\cite{Green:1998by,Basu:2008cf}.

\subsection{Information from the tree level 4 graviton amplitude}

We shall find it useful in our analysis to consider the contribution to the $D^{6n} \mathcal{R}^4$ interactions coming from the tree level (genus zero) 4 graviton amplitude in type II string theory, and so we briefly mention the results.

The genus zero 4 graviton amplitude is given by
\be \label{tree} l_s^{-8}A^{(4)} (s,t,u)= \frac{64}{{l_s}^6 stu} e^{-2\phi}\frac{\Gamma (1-l_s^2 s/4)\Gamma(1-l_s^2 t/4)\Gamma(1-l_s^2 u/4)}{\Gamma(1+l_s^2 s/4)\Gamma(1+l_s^2 t/4)\Gamma(1+l_s^2 u/4)} t_8 t_8 R^4,\ee
where $s,t$ and $u$ are the Mandelstam variables. The low momentum expansion of \C{tree} yields the various terms in the type II effective action at tree level. Using the identity
\be {\rm ln} ~\Gamma (1-z) = \gamma z + \sum_{n=2}^\infty \frac{\zeta (n)}{n} z^n,\ee 
where $\gamma$ is the Euler--Mascheroni constant, we can easily write down the terms relevant for us. Among them, we write down some of the terms at low orders in the $l_s$ expansion, to see the structure that arises. We shall find the low momentum expansion of the genus 0 amplitude to be useful in our later analysis. 

In obtaining the expressions below, we make use of the relation~\cite{Green:1999pv}
\be s^k + t^k + u^k = k\sum_{2p+3q=k} \frac{(p+q-1)!}{p!q!} \Big(\frac{s^2 + t^2 + u^2}{2} \Big)^p \Big( \frac{s^3 + u^3 + t^3}{3}\Big)^q ,\ee
where $s+t+u=0$.

Beyond the Einstein--Hilbert action, the leading contact interaction is the $\mathcal{R}^4$ term (hence $n=0$) given by
\be \label{n0} 2\zeta (3)l_s^{8}t_8 t_8 R^4,\ee
which is obtained from \C{tree}. The next term, for $n=1$ is the 
$D^6 \mathcal{R}^4$ interaction given by
\be \label{n1}\frac{2}{3} \zeta (3)^2 (l_s/2)^6(s^3 + t^3 + u^3) l_s^8t_8 t_8 R^4.\ee
For $n=2$, the $D^{12} \mathcal{R}^4$ interaction is given by
\be \label{n2}(l_s/2)^{12}\Big[ \frac{4}{27}\zeta (3)^3 (s^3 + t^3 + u^3)^2 +2 \zeta (9)\Big(\frac{1}{8} (s^2 +  t^2 + u^2)^3 + \frac{1}{27} (s^3 + t^3 + u^3)^2 \Big) \Big]l_s^8 t_8 t_8 R^4.\ee

Thus for $n=2$, there are two distinct types of spacetime structures, given by $(l_s/2)^{12} (s^3 + t^3 + u^3)^2$ and $(l_s/2)^{12}(s^2 + t^2 + u^2)^3$, with coefficients $4\zeta (3)^3 /27 +2\zeta (9)/27$ and $\zeta (9)/4$ respectively.

For $n=3$, the $D^{18} \mathcal{R}^4$ interaction is given by
\bea \label{n3} &&(l_s/2)^{18} \Big[ \frac{1}{3} \zeta (3) \zeta (9) (s^3 + t^3 + u^3)\Big( \frac{1}{2} (s^2 + t^2 + u^2)^3 + \frac{4}{27} (s^3 + t^3 + u^3)^2 \Big) \non \\ &&+ \frac{2}{81} \zeta (3)^4(s^3 + t^3 + u^3)^3 + \frac{1}{6} \zeta (5) \zeta (7) (s^2 + t^2 + u^2)^3 (s^3 + t^3 + u^3)\Big] l_s^8t_8 t_8 R^4.\eea
Once again there are two distinct types of spacetime structures, given by $(l_s/2)^{18}(s^3 + t^3 + u^3)^3$ and $(l_s/2)^{18} (s^3 + t^3 + u^3) (s^2 + t^2 + u^2)^3$ with coefficients $2\zeta(3)^4 /81+ 4 \zeta (3) \zeta (9)/81$ and $\zeta (5) \zeta (7)/6 +\zeta (3)\zeta (9)/6$ respectively.  

Finally, for $n=4$, the $D^{24} \mathcal{R}^4$ interaction is given by
\bea \label{24} &&(l_s/2)^{24} \Big[ \frac{1}{32} \zeta (15) (s^2 + t^2 + u^2)^6 +\frac{2}{81}\Big( \frac{2}{15} \zeta (3)^5 + \frac{2}{3} \zeta(3)^2 \zeta (9) + \frac{1}{5} \zeta (15) \Big)(s^3 +  t^3 + u^3)^4 \non \\ &&+\Big( \frac{1}{18} \zeta(3)^2 \zeta (9) + \frac{1}{9} \zeta (3) \zeta (5) \zeta (7)  +\frac{1}{54} \zeta(5)^3 + \frac{5}{54} \zeta (15)\Big)(s^2 + t^2 + u^2)^3 (s^3 + t^3 + u^3)^2 \Big] \non \\ &&\times l_s^8 t_8t_8 R^4,\eea
which has three distinct spacetime structures with the coefficients mentioned above.

This pattern continues for all $n$. 
For arbitrary $n$, there is always a $D^{6n} \mathcal{R}^4$ interaction of the specific spacetime form 
\be \label{univ} \frac{2^{n+1}}{3^n (n+1)!} \Big[\zeta (3)^{n+1} +\ldots\Big](l_s/2)^{6n}(s^3 + t^3 + u^3)^n l_s^8 t_8 t_8 R^4\ee
which we refer to as the universal interaction, which will be very useful for our purposes. The other distinct spacetime interactions at this order in the $l_s$ expansion all have coefficients of transcendentality $3(n+1)$ and involve Riemann zeta functions each having odd transcendentality greater than 1. The universal interaction \C{univ} is the only one which has the maximum number of zeta functions consistent with this, and all are $\zeta (3)$. It will be clear from our analysis in what sense these various interactions, at the same order in the $l_s$ expansion, are distinct.

Our primary aim is not to analyze genus zero interactions in string theory, but to consider the nature of the $D^{6n} \mathcal{R}^4$ interactions in the full effective action in M theory. On compactifying on $T^2$ and taking the volume of the two torus to zero to obtain type IIB string theory in 10 flat dimensions, the moduli dependent couplings of the $D^{6n} \mathcal{R}^4$ interactions receive perturbative contributions from various string genera, as well as non--perturbative contributions from D--instantons. In the Einstein frame, the coefficients of the $D^{6n} \mathcal{R}^4$ interactions are $SL(2,\mathbb{Z})$ invariant modular forms on the fundamental domain of $SL(2,\mathbb{Z})$. It is of interest to determine these modular forms, whose genus zero contributions in the string frame must match the contributions obtained from \C{tree}. The genus $n+1$ contributions to these modular forms are what we want to calculate to possibly obtain the nature of the interaction in M theory.   

Among the modular forms relevant for our purposes, the exact expressions are known only for $n=0$ and $1$, as the $\mathcal{R}^4$ and $D^6 \mathcal{R}^4$ interactions are $1/2$ BPS and $1/8$ BPS respectively. For $n\geq 2$, the $D^{6n} \mathcal{R}^4$ interactions in the type II theory (and consequently in M theory) are non--BPS and do not satisfy simple non--renormalization theorems like their BPS counterparts.

\subsection{The $\mathcal{R}^4$ and $D^6 \mathcal{R}^4$ interactions in string and M theory}

In the string frame in type IIB string theory, for $n=0$, the genus 0 amplitude \C{n0} is the leading contribution to
\be \label{f0} l_s^{-2} \int d^{10} x \sqrt{-g} e^{-\phi_B/2} f^{(0)} (\tau,\bar\tau)\mathcal{R}^4,\ee  
where $f^{(0)} (\tau,\bar\tau)$ is given by the Eisenstein series $E_{3/2}(\tau,\bar\tau)$~\cite{Green:1997tv} which satisfies
\be \label{E32} 4 \tau_2^2 \frac{\p^2}{\p \tau \p \bar\tau} E_{3/2} = \frac{3}{4} E_{3/2}\ee
on the fundamental domain of $SL(2,\mathbb{Z})$. 
Now $E_{3/2}$ has only genus 0 and 1 perturbative contributions given by
\be \label{n0more} f^{(0)} (\tau,\bar\tau) = E_{3/2} (\tau,\bar\tau) = 2\zeta (3) \tau_2^{3/2} +4 \zeta (2) \tau_2^{-1/2} +\ldots.\ee 
Here $\tau= c_0^B + i e^{-\phi_B}$ is the complex type IIB modulus where $c_0^B$ is the R--R pseudoscalar and $\phi_B$ is the type IIB dilaton, and the terms neglected in \C{n0more} involve D--instanton contributions, which we shall also drop in all the discussions below. Thus from \C{actM} and \C{actIIA}, ignoring an overall factor, it follows that the $\mathcal{R}^4$ interaction in M theory is given by
\be \label{act0}l_{11}^{-3} \zeta (2) \int d^{11} x \sqrt{-G} \mathcal{R}^4,\ee
thus the coefficient is fixed by the 1 loop $\mathcal{R}^4$ term in \C{n0more}. In the various expressions for the interactions in M theory, we use the metric $G_{MN}$ from \C{metric}.

For $n=1$, in the string frame the genus 0 contribution \C{n1} is the leading contribution to
\be \label{f6}l_s^4 \int d^{10} x \sqrt{-g} e^{\phi_B} f^{(6)} (\tau,\bar\tau) D^6 \mathcal{R}^4 ,\ee
where $f^{(6)} (\tau,\bar\tau)$ satisfies the Poisson equation~\cite{Green:2005ba}
\be \label{n1more} 4 \tau_2^2 \frac{\p^2}{\p \tau \p \bar\tau} f^{(6)} = 12 f^{(6)} - E_{3/2}^2.\ee
From \C{n1more}, if follows that $f^{(6)} (\tau,\bar\tau)$ receives perturbative contributions only upto genus 3 given by
\be \label{2loop}f^{(6)} (\tau,\bar\tau) = \frac{2}{3}\zeta (3)^2 \tau_2^3 + \frac{4}{3}\zeta (2) \zeta (3) \tau_2 + \frac{8}{5} \zeta (2)^2 \tau_2^{-1}+ \frac{32}{945}\zeta (2)^3 \tau_2^{-3} +\ldots.\ee
While the genus 0, 1 and 2 contributions follow from the source term in \C{n1more}, calculating the genus 3 contribution requires more effort as the coefficient is automatically a solution of the homogeneous equation in \C{n1more}. This analysis involves multiplying \C{n1more} by $E_4$ which satisfies
\be 4 \tau_2^2 \frac{\p^2}{\p \tau \p \bar\tau} E_4 = 12 E_4,\ee
 and then integrating the equation over the fundamental domain of $SL(2,\mathbb{Z})$ to extract the relevant coefficient. 
  
Again from \C{actM} and \C{actIIA}, ignoring an overall factor, it follows that the $D^6 \mathcal{R}^4$ interaction in M theory is given by
\be \label{twoloop} l_{11}^3\zeta (2)^2 \int d^{11}x \sqrt{-G} D^6 \mathcal{R}^4 ,\ee 
where the coefficient is fixed by the 2 loop $D^6 \mathcal{R}^4$ term in \C{2loop}.

Note that the $\mathcal{R}^4$ and $D^6 \mathcal{R}^4$ interactions satisfy non--renormalization theorems as a consequence of which they receive only a finite number of perturbative contributions. This is because these interactions are BPS.  

\subsection{Leading contributions to the $D^{12}\mathcal{R}^4$, $D^{18} \mathcal{R}^4$ and $D^{24} \mathcal{R}^4$ interactions in the type II theory}

For $n \geq 2$, no explicit expressions are available for the non--BPS $D^{6n} \mathcal{R}^4$ interactions. However, it follows from the structure of the various coefficients that arise from \C{tree} that the different Riemann zeta functions are the genus 0 approximations to different modular forms that are the couplings of the various interactions. Thus for $n=2$, from \C{n2} we see that the 2 distinct spacetime interactions must have two distinct modular forms as their coefficients given by $f_1^{(12)}$ and $f_2^{(12)}$, where 
\be \label{univ12}f_1^{(12)} = \frac{2}{27}\Big(2\zeta(3)^3 +  \zeta (9) \Big)\tau_2^{9/2} +\ldots ,\ee 
and 
\be \label{nonuniv12}f_2^{(12)} =  \frac{1}{4}\zeta (9) \tau_2^{9/2} +\ldots\ee 
in the Einstein frame. Similarly for $n=3$, given \C{n3} it follows that there are two distinct modular forms $f_1^{(18)}$ and $f_2^{(18)}$ whose weak coupling expansions are given by
\be \label{univ18}f_1^{(18)} = \frac{2}{81}\Big( \zeta (3)^4 + 2 \zeta (3) \zeta (9) \Big)\tau_2^6+\ldots,\ee and
\be \label{nonuniv18}f_2^{(18)} = \frac{1}{6}\Big(  \zeta (5)\zeta (7) + \zeta (3) \zeta (9)\Big)\tau_2^6 +\ldots \ee 
in the Einstein frame. 

Finally, for $n=4$  from \C{24} there must be three distinct modular forms $f^{(24)}_1, f_2^{(24)}$ and $f_3^{(24)}$ whose weak coupling expansions are
\bea \label{24form}f_1^{(24)} &=& \frac{2}{81} \Big( \frac{2}{15} \zeta (3)^5 + \frac{2}{3} \zeta (3)^2 \zeta (9) +\frac{1}{5} \zeta (15) \Big), \non \\ f_2^{(24)} &=& \frac{1}{32} \zeta (15), \non \\ f_3^{(24)} &=& \frac{1}{9} \Big( \frac{1}{2} \zeta (3)^2 \zeta (9) + \zeta (3) \zeta (5) \zeta (7) + \frac{1}{6} \zeta (5)^3 + \frac{5}{6} \zeta (15)\Big). \eea

These are the cases we shall describe in some detail in the sections to come. This pattern continues for all $n$, and the number of independent modular forms increases rapidly as $n$ increases. 

\section{Constraints on the type IIB effective action from supersymmetry and S--duality, and M theory interactions}

Let us now analyze the results mentioned above using the constraints imposed on the effective action by supersymmetry and S--duality~\cite{Green:1998by,Basu:2008cf}. This will also help us generalize the arguments for $n \geq 2$.

The main output obtained by imposing the constraints of supersymmetry and S--duality for a certain class of interactions in the type IIB effective action is that the moduli dependent coefficients of these interactions satisfy first order differential equations on moduli space. This holds to all orders in the $l_s$ expansion. We shall simply write down very schematically the structure of these equations which is good enough for our purposes. We refer the reader to~\cite{Basu:2008cf} for the details. These equations were derived for the $\hat{G}^{2k} \lambda^{16}$ interactions for all $k \geq 0$, which should lie in the same supermultiplet as the $D^{2k} \mathcal{R}^4$ interactions\footnote{Here $\hat{G}_{\mu\nu\lambda}$ is the supercovariant 3--form field strength of type IIB supergravity given by~\cite{Schwarz:1983qr}
\be \hat{G}_{\mu\nu\lambda} = G_{\mu\nu\lambda} -3 \bar\psi_{[\mu} \g_{\nu\lambda]} \lambda - 6i \bar\psi^*_{[\mu} \g_\nu \psi_{\lambda]},\ee
where $G_{\mu\nu\lambda}$, $\psi_\mu$ and $\lambda$ are the $SL(2,\mathbb{R})$ invariant 3--form field strength, the gravitino and the dilatino respectively.

The complete structure of these non--BPS multiplets is not known, so it not clear exactly which interactions lie in the same multiplet. However the fact that the couplings satisfy \C{main} for all the interactions follows from the general structure of supersymmetry and the Noether construction. It is simply that the maximally fermionic terms are easiest to analyze, as discussed in~\cite{Basu:2008cf}.}. Then the structure of these equations follows for all the moduli dependent couplings of the various interactions in the supermultiplet by supersymmetry, though it is difficult to implement it at the non--linear level.      

These first order equations for the various interactions in the supermultiplet are given by
\bea \label{fo}D f \sim f' +\sum_i g_i h_i , \quad
\bar{D} f' \sim f + \sum_i k_i l_i.\eea
In \C{fo}, $D$ and $\bar{D}$ are appropriate $SL(2,\mathbb{Z})$ modular covariant derivatives and $f,f',g_i,h_i, k_i$ and $l_i$ are various couplings where $f$ and $f'$ are at the same order in the $l_s$ expansion, and the source terms involve $g_i,h_i,k_i$ and $l_i$  which are at lower orders in the $l_s$ expansion.
Iterating them, we get that the coupling $f$ of every interaction must satisfy the Poisson equation
\be \label{main}4\tau_2^2 \frac{\p^2}{\p \tau \p \bar\tau} f \sim f+ \sum_i r_i s_i + \sum_i m_i n_i p_i, \ee
on the fundamental domain of $SL(2,\mathbb{Z})$, where $r_i,s_i,m_i,n_i$ and $p_i$ are couplings of interactions at lower orders in the derivative expansion. Note that the source terms can be at most cubic in the coefficient functions. Thus given appropriate boundary conditions, one can solve for $f$ in \C{main} if one knows the modular forms that arise as source terms at lower orders in the $l_s$ expansion. Hence this system of equations can be solved recursively.  We now consider the consequences of this for the $D^{6n} \mathcal{R}^4$ interactions.

\subsection{Implementing the constraints}

To start with, we use the relation
\be \label{inv}\delta S= 0,\ee
which is the statement of invariance of the effective action under supersymmetry transformations upto total derivatives. We now expand the effective action and the supersymmetry transformations in powers of $l_s$ given by
\bea \label{sum}S &=& S^{(0)} + \sum_{m=3}^\infty l_s^{2m} S^{(m)} , \non \\ \delta &=& \delta^{(0)} + \sum_{m=3}^\infty l_s^{2m} \delta^{(m)}, \eea
where $S^{(0)}$ is the supergravity action, and $\delta^{(0)}$ denotes the supersymmetry transformations which leave the supergravity Lagrangian invariant upto a total derivative\footnote{Due to the absence of a covariant action without auxiliary fields of type IIB supergravity because of the self--dual 5 form, what is actually meant is the invariance of the equations of motion.}. The correction to $S^{(0)}$ starts at $O(l_s^6)$ because the $\mathcal{R}^2$ and $\mathcal{R}^3$ supermultiplets vanish due to maximal supersymmetry, and so the first correction to Einstein gravity is given by the $\mathcal{R}^4$ interaction. Thus the corresponding terms in $\delta$ vanish as well. 

Now starting from \C{inv} and \C{sum}, one can obtain results imposing the invariance of the action under supersymmetry transformations order by order in the $l_s$ expansion. The $D^{6n} \mathcal{R}^4$ interaction is one of the terms in $S^{(3n +3)}$.
Thus we are interested in the constraints imposed by \C{inv} at $O(l_s^{6n+6})$ for $n \geq 0$.

\subsection{The $\mathcal{R}^4$ and $D^6 \mathcal{R}^4$ interactions}

For $n=0$, from \C{inv} we get the relation
\be \delta^{(0)} S^{(3)} +\delta^{(3)} S^{(0)} =0,\ee
which leads to the coefficient of the $\mathcal{R}^4$ interaction given in \C{n0more}, which satisfies Laplace equation on moduli space \C{E32}.
The coefficient $3/4$ in \C{E32} is fixed given that the genus 0 contribution $\sim \tau_2^{3/2}$.

For $n=1$, from \C{inv} we get that
\be \delta^{(0)} S^{(6)} +\delta^{(6)} S^{(0)} +\delta^{(3)} S^{(3)}=0,\ee
which leads to the coefficient of the $D^6 \mathcal{R}^4$ interaction given in \C{n1more}. The $\delta^{(3)} S^{(3)}$ term leads to the source term $E_{3/2}^2$ in the Poisson equation~\cite{Basu:2008cf}. Now the $D^6 \mathcal{R}^4$ interaction receives perturbative contributions only upto genus 3~\cite{Berkovits:2006vc}, while the source terms in \C{n1more} yield contributions only upto genus 2. Hence the coefficient of the homogeneous part of \C{n1more} is fixed to be $12$ given that the genus 3 contribution $\sim \tau_2^{-3}$.

Thus we see that for $n=0$ and $n=1$, the equations for the moduli dependent coefficients are completely determined by considerations of supersymmetry and S--duality.

\subsection{The $D^{12} \mathcal{R}^4$ interaction}

Let us now analyze the $n=2$ case. From \C{inv}, we get that
\be \label{eqn1}\delta^{(0)} S^{(9)} + \delta^{(9)} S^{(0)} + \delta^{(3)} S^{(6)} + \delta^{(6)} S^{(3)} + \delta^{(4)} S^{(5)} + \delta^{(5)} S^{(4)}= 0.\ee
Thus the $\delta^{(3)} S^{(6)}, \delta^{(6)} S^{(3)}, \delta^{(4)} S^{(5)}$ and $\delta^{(5)} S^{(4)}$ terms in \C{eqn1} provide source terms to the homogeneous equation obtained from $\delta^{(0)} S^{(9)}$ and $\delta^{(9)} S^{(0)}$, and so the moduli dependent couplings satisfy Poisson equations.

For $n=2$, from \C{n2} and the subsequent arguments, it follows that one has to determine the two $SL(2,\mathbb{Z})$ invariant modular forms $f_1^{(12)}$ and $f_2^{(12)}$ whose weak coupling expansions are given by \C{univ12} and \C{nonuniv12} respectively. A detailed analysis of these kind of non--BPS interactions for $n=2$ or for higher $n$ has not been done. However, we can obtain some constraints on $f_i^{(12)}$ based on supersymmetry and S--duality, which we now describe. This also leads to constraints for the $D^{12} \mathcal{R}^4$ term in the M theory effective action.

First consider the source terms that contribute to the Poisson equation for $n=2$, and which arise from $\delta^{(3)} S^{(6)}$ and $\delta^{(6)} S^{(3)}$. The contributions from $S^{(3)}$ and $S^{(6)}$ are given by $f^{(0)}= E_{3/2}$ and $f^{(6)}$ respectively, where the expressions are  given by the solutions to \C{E32} and \C{n1more} respectively. The corrected supervariations $\delta^{(3)}$ ($\delta^{(6)}$) are then proportional to the modular form of the coupling in $S^{(6)}$ ($S^{(3)}$). 
Thus the total source term contribution to the Poisson equation is a linear combination of $E_{3/2} f^{(6)}$ and $E_{3/2}^3$. 

Now consider the source terms which arise from $\delta^{(4)} S^{(5)}$ and $\delta^{(5)} S^{(4)}$.
The contribution from $S^{(5)}$ is given by
\be E_{5/2} (\tau,\bar\tau) = 2\zeta (5) \tau_2^{5/2} + \frac{8}{3} \zeta (4) \tau_2^{-3/2} +\ldots,\ee 
where $E_{5/2}$ is the coupling of the $1/4$ BPS interaction $D^4 \mathcal{R}^4$, given in the string frame by~\cite{Green:1999pu,Sinha:2002zr}
\be l_s^2 \int d^{10}x \sqrt{-g} e^{\phi_B/2}E_{5/2} (\tau,\bar\tau) D^4 \mathcal{R}^4.\ee
That this coupling satisfies the Laplace equation
\be 4 \tau_2 \frac{\p^2}{\p \tau \p \bar\tau} E_{5/2} = \frac{15}{4} E_{5/2}\ee
follows from the constraints of supersymmetry, while the factor $15/4$ is fixed because the genus 0 contribution $\sim \tau_2^{5/2}$ which follows from \C{tree}. 

The contribution from $S^{(4)}$ is not so obvious, and has been discussed in~\cite{Basu:2013goa} in the context of the source terms for the $\mathcal{R}^8$ interaction. To see this, note that the genus one 4 graviton amplitude in the type II theory has a term which is non--analytic in the external momenta, and is schematically given by $\zeta (2) s {\rm ln} (-l_s^2 s) \mathcal{R}^4$~\cite{Green:2006gt,Green:2008uj} in the string frame, appropriately symmetrized in $s,t$ and $u$. In the Einstein frame, this leads to an interaction proportional to $\zeta (2){\rm ln} \tau_2(s+t+u)\mathcal{R}^4$ in the effective action in $S^{(4)}$, which vanishes on--shell using $s+t+u=0$. However, this interaction would survive in any off--shell formulation of the theory, and would give a contribution from $S^{(4)}$ of the form $Y(\tau,\bar\tau)$, where
\be \label{Y} Y(\tau,\bar\tau) = \zeta (2){\rm ln}\tau_2 + \ldots\ee
is the $SL(2,\mathbb{Z})$ invariant completion of $\zeta (2) {\rm ln}\tau_2$ which has no other perturbative contributions. 

We keep this interaction in our supersymmetry analysis, because the equations derived for the moduli dependent couplings are exact equations, and should be valid even off--shell. They make no reference to the specific details of the interactions and give the complete non--perturbative answer. Again, the corrected supervariations $\delta^{(4)}$ ($\delta^{(5)}$) are then proportional to the modular form of the interaction in $S^{(5)}$ ($S^{(4)}$). Thus the contribution of these source terms to the Poisson equation is given by $E_{5/2} Y$. 

Thus even though the $D^{12} \mathcal{R}^4$ interaction is non--BPS, the source terms that arise in the Poisson equations that determine its coupling are BPS.
Given the structure of the source terms,
let us consider the modular form $f_1^{(12)}$ for the universal interaction \C{univ12} for $n=2$ which has a linear combination of $\zeta(3)^3$ and $\zeta (9)$ as its tree level coefficient. Clearly, 
\be \label{eqn2} f_1^{(12)}(\tau,\bar\tau) = \sum_i f_{1,i}^{(12)} (\tau,\bar\tau),\ee
where $f_{1,i}^{(12)}$ satisfies
\be \label{eqn3} 4 \tau_2^2 \frac{\p^2}{\p \tau \p \bar\tau} f_{1,i}^{(12)} = \lambda_i f_{1,i}  - \mu_i E_{3/2} f^{(6)} - \nu_i E_{3/2}^3 - \eta_iE_{5/2} Y\ee
for every $i$. 

The structure of the equation satisfied for each $f_{1,i}$ given in \C{eqn3} follows from our general discussion above regarding the constraints imposed by supersymmetry, as explained in~\cite{Basu:2008cf}. The fact that $f_1^{(12)}$ must split into a sum of modular forms as given in \C{eqn2} follows because the source terms yield contributions only upto genus 4. However the non--BPS interaction $D^{12} \mathcal{R}^4$ receives contributions from higher genera as well, which must be accounted for by the homogeneous terms in the various equations \C{eqn3}. Thus, for every $\lambda_i$ that is consistent with string perturbation theory, we have to look at one equation for $f_{1,i}^{(12)}$ in \C{eqn3}.

For example, the genus 5 contribution $\sim \tau_2^{-11/2}$ will be contained in the modular form $g_1$ which satisfies
\be 4 \tau_2^2 \frac{\p^2}{\p \tau \p \bar\tau} g_1 = \frac{143}{4} g_1  -  \alpha E_{3/2} f^{(6)} - \beta E_{3/2}^3 -\gamma  E_{5/2} Y,\ee 
while the genus 6 contribution $\sim \tau_2^{-15/2}$ will be contained in the modular form $g_2$ which satisfies
\be \label{6loop} 4 \tau_2^2 \frac{\p^2}{\p \tau \p \bar\tau} g_2 = \frac{255}{4} g_2  -  \eta E_{3/2} f^{(6)} -\omega E_{3/2}^3 -\sigma E_{5/2}Y,\ee
where $\alpha,\beta,\gamma,\eta,\omega$ and $\s$ are numerical factors of vanishing transcendentality.
This splitting occurs for all non-BPS  interactions from $D^8 \mathcal{R}^4$ onwards in 10 dimensions~\cite{Basu:2008cf}, and has also been observed in 9 dimensions in the calculation of the two loop 4 graviton amplitude in maximal supergravity~\cite{Green:2008bf}.  

To obtain the nature of the M theory interaction for $n=2$ that results from $f_1^{(12)}$, we need to isolate the 3 loop type IIB amplitude, and so we consider \C{eqn3} in detail, as the total contribution is given by \C{eqn2}. The perturbative contributions from the various source terms are given by
\bea \label{manycont}
E_{3/2} f^{(6)} &=& \frac{4}{3}\zeta(3)^3 \tau_2^{9/2} + \frac{16}{3}\zeta (2) \zeta (3)^2 \tau_2^{5/2} + \frac{128}{15}\zeta(2)^2 \zeta (3) \tau_2^{1/2} \non \\&&+ \frac{32}{5}\zeta(2)^3 \Big(1+ \frac{2}{189}\zeta (3)\Big) \tau_2^{-3/2} + \frac{128}{945}\zeta(2)^4 \tau_2^{-7/2}, \non \\ E_{3/2}^3 &=& 8 \zeta (3)^3 \tau_2^{9/2} + 48 \zeta (2) \zeta (3)^2 \tau_2^{5/2} + 96 \zeta (2)^2 \zeta (3) \tau_2^{1/2} + 64 \zeta (2)^3 \tau_2^{-3/2}, \non \\ E_{5/2} Y &=& 2\zeta (2)\zeta(5) \tau_2^{5/2}{\rm ln} \tau_2 + \frac{8}{3}\zeta (2)\zeta (4) \tau_2^{-3/2} {\rm ln} \tau_2.\eea
Using these source term contributions and with appropriate boundary conditions, we can solve \C{eqn3}. To see the structure that arises with a particular example, we solve for the case where the equation is given by \C{6loop}. This equation must be there among the set of equations \C{eqn3} as the genus 6 contribution to $D^{12} \mathcal{R}^4$ is non--vanishing and is given by the type IIA genus 6 amplitude, which is completely determined by 1 loop supergravity\footnote{The genus $n$ type IIA $D^{2n} t_8t_8R^4$ amplitude is given by the expression~\cite{Green:1999pu} 
\bea \label{genusn}\frac{8\pi^2}{n!} \Gamma (n-1) \zeta (2n-2) e^{2(n-1)\phi_A} l_s^{2n} \mathcal{W}^n \mathcal{R}^4,\eea
where $\mathcal{W}^n = O(s^n)$, and is symmetric in $s,t$ and $u$.
Thus the genus 6 type IIA $D^{12} t_8t_8R^4$ amplitude is equal to  
\bea \frac{1820}{691} \zeta (12) e^{10\phi_A} l_s^{12} \mathcal{W}^6 \mathcal{R}^4,\eea
 where we have used 
\be \zeta (2) = \frac{\pi^2}{6}, \quad \zeta (10) = \frac{\pi^{10}}{93555}, \quad \zeta (12)=\frac{691\pi^{12}}{638512875}.\ee
This is a calculation in the strongly coupled IIA theory, and we assume the validity of its transcendentality at weak coupling like the other genus amplitudes. 
}. Recall that we are looking at the $t_8 t_8 R^4$ part of the amplitude, and so the perturbative part of the amplitude is the same for both the type II theories. Solving \C{6loop}, we get that
\bea \label{ansloop}g_2 (\tau,\bar\tau)&=& \frac{1}{12} \Big( \frac{\eta}{3} + 2\omega\Big)\zeta (3)^3 \tau_2^{9/2} + \frac{4}{15} \Big( \frac{\eta}{3} + 3\omega\Big)\zeta (2) \zeta (3)^2 \tau_2^{5/2} +\frac{1}{2} \Big( \frac{4\eta}{15} + 3\omega\Big)\zeta (2)^2 \zeta (3) \tau_2^{1/2} \non \\&& + \frac{8}{15} \Big( \frac{1}{5}  \Big\{1+\frac{2}{189}\zeta (3)\Big\}\eta +2\omega\Big)\zeta (2)^3  \tau_2^{-3/2} + \frac{8\eta}{2835} \zeta (2)^4 \tau_2^{-7/2} + \zeta (12) \tau_2^{-15/2} \non \\&&
+ \frac{\s}{30} \zeta (2) \zeta (5) \tau_2^{5/2} {\rm ln}\tau_2 + \frac{2\s}{45} \zeta (2)\zeta (4) \tau_2^{-3/2} {\rm ln}\tau_2\non \\ &&+\frac{\s}{450} \zeta (2)\zeta (5) \tau_2^{5/2} - \frac{2\s}{675} \zeta (2)\zeta (4) \tau_2^{-3/2}.\eea

We have neglected the solution $\sim \tau_2^{17/2}$ to the homogeneous equation \C{6loop} as it does not have the correct $\tau_2$ dependence to be a string amplitude. Also we have set the genus 6 coefficient to be $\zeta (12)$ by absorbing an overall constant in the definition of $g_2$.

\subsubsection{The analysis for generic $\lambda_i$}

The structure that has been obtained in \C{ansloop} for $\lambda_i = 255/4$ easily generalizes for generic $\lambda_i$ in \C{eqn3} (specific choices of $\lambda_i$ have to be analyzed separately, which we shall do later). To be consistent with string perturbation theory, $\lambda_i$ must be of the form
\be \label{deflam}\lambda_i = s_i (s_i -1),\ee 
so that the homogeneous equation of \C{eqn3} has the perturbative solutions $\tau_2^{s_i}$ and $\tau_2^{1-s_i }$. If the power behaviour of any of these solutions is inconsistent with its interpretation as a string amplitude, then its coefficient vanishes (for example, the coefficient of the $\tau_2^{17/2}$ term was set to 0 in \C{ansloop}). Thus, for generic $\lambda_i$ in \C{eqn3} given by \C{deflam}, the solution is given by

\bea \label{ans6loop}f^{(12)}_{1,i} (\tau,\bar\tau)&\sim& \alpha_i\zeta (3)^3 \tau_2^{9/2} + \beta_i \zeta (2) \zeta (3)^2 \tau_2^{5/2} +\gamma_i \zeta (2)^2 \zeta (3) \tau_2^{1/2} \non \\ &&+ \Big( \epsilon_i + \mu_i \zeta (3)\Big)\zeta (2)^3  \tau_2^{-3/2} + \mu_i \zeta (2)^4 \tau_2^{-7/2} \non \\ && +c_{1,i}\tau_2^{s_i} + c_{2,i} \tau_2^{1-s_i}\non \\&&
+ \eta_i \zeta (2) \Big( 2 \zeta (5) \tau_2^{5/2} + \frac{8}{3} \zeta (4) \tau_2^{-3/2}\Big) {\rm ln}\tau_2\non \\ &&+\eta_i  \zeta (2)\zeta (5) \tau_2^{5/2} + \eta_i \zeta (2)\zeta (4) \tau_2^{-3/2},\eea
where $\alpha_i, \beta_i, \gamma_i$ and $\epsilon_i$ are linear combinations of $\mu_i$ and $\nu_i$ with $s_i$ dependent coefficients. In \C{ans6loop} we have not written down the various numerical factors as they are not relevant for our purpose.   
The structure obtained in \C{ans6loop} is also true for $f^{(12)}_{2,i}$.

Let us now consider the various terms in \C{ans6loop}.
The first two lines of \C{ans6loop} give genus 0 to 4 contributions to the modular form $f^{(12)}_{1,i}$. The genus 0 contribution matches the $\zeta (3)^3$ part of \C{univ12} , while the structure of the genus 1 contribution agrees with the string calculation~\cite{Green:2008uj}. The third line has the two solutions of the homogeneous equation \C{eqn3}. The fourth line gives the non--analytic contributions, which yield non--local interactions of the schematic form $\zeta (2)\zeta (5){\rm ln}(-l_s^2 s) D^{12} \mathcal{R}^4$ and $\zeta (2)\zeta (4) e^{4\phi_B}{\rm ln} (-l_s^2 s) D^{12} \mathcal{R}^4$ in the type IIB effective action in the string frame, at genus 1 and 3 respectively, appropriately symmetrized in $s,t$ and $u$. 
The genus 1 non--analytic contribution indeed has precisely this structure~\cite{Green:2008uj}, hence justifying the presence of this source term involving $Y$. The other higher genus amplitudes have not been calculated in string theory.

Now let us consider the  analytic parts of the genus 3 and 4 amplitudes obtained from \C{ans6loop}, and compare with the 1 and 2 loop maximal supergravity calculations in~\cite{Green:2008bf}. As explained in~\cite{Green:2008bf}, the supergravity calculation is valid in the limit when the dimensionlesss volume of the compact manifold is much greater than 1, which translates into strong coupling in the type IIA theory. The supergravity calculation includes the terms (dropping overall numerical factors)
\be \label{sugra} \zeta(2)^3 \zeta (3) e^{4\phi_A} + \zeta (2)^4 e^{6\phi_A} +\zeta(2)^6  e^{10\phi_A}.\ee
Hence the transcendentality of the coefficients of the strong coupling expansion with this dilaton dependence precisely matches the transcendentality at weak coupling in \C{ans6loop} after converting to the string frame. This non--trivial agreement of the transcendentality of the coefficients at weak and strong coupling is the basis for our assumption in this work, that this feature generalizes at least for small $n$. We expect this to be a consequence of maximal supersymmetry, and even non--BPS interactions are tightly constrained. Note that the $\zeta (2)^3 \tau_2^{-3/2}$ part of the genus 3 amplitude in \C{ans6loop} is not obtained from \C{sugra}. However, given the matching of the transcendentality of the $\zeta (2)^3 \zeta (3) \tau_2^{-3/2}$ part, it is natural to expect that the $\zeta (2)^3 \tau_2^{-3/2}$ term also survives at strong coupling. It would be interesting to see if higher loop quantum supergravity can reproduce this contribution.

Finally, let us consider the two terms in the last line of \C{ans6loop}. These give genus 1 and 3 contributions to $f^{(12)}_{1,i}$ with coefficients $\zeta (2) \zeta (5)$ and $\zeta (2) \zeta (4)$ respectively. Both these contributions are proportional to $\eta_i$, and hence come from the non--analytic part of the source terms.  While the existence of these terms for each individual equation in \C{eqn3} is crucial for S-duality,   
we argue that these contributions can be neglected in the final answer. For the genus 1 contribution, our argument follows from the string calculation of~\cite{Green:2008uj}. 

 The non--analytic terms that appear in \C{ans6loop} lead to  terms logarithmic in the Mandelstam variables in the string frame in the low energy effective action, which are infra--red divergent. One always has an ambiguity to choose the scale of the logarithm to shift the coefficient of the infra--red finite term. Thus the shift of the scale of the logarithm $\mu_1$ from the infra--red divergent $\zeta (2) \zeta (5) s^6 {\rm ln}(-\mu_1 l_s^2 s) \mathcal{R}^4$ term adds a contribution proportional to $\zeta (2) \zeta (5) s^6 \mathcal{R}^4$ to the infra--red finite genus 1 coefficient, while the shift of the scale $\mu_3$ from the infra--red divergent $\zeta (2)\zeta (4) e^{4\phi_B} s^6{\rm ln} (-\mu_3 l_s^2 s)  \mathcal{R}^4$ term adds a contribution proportional to $\zeta (2)^3  e^{4\phi_B}s^6 \mathcal{R}^4$ to the infra--red finite genus 3 coefficient. For the genus 1 contribution, adding all these contributions together that arise from the various equations in \C{eqn2}, we simply set these terms to 0 by appropriately choosing the scale of the logarithm.  In fact in string theory, this can be done unambiguously. The genus 1 infra--red finite piece is proportional to $\zeta (2) \zeta (3)^2 \tau_2^{5/2}$ while the infra--red divergent part is proportional to $\zeta (2) \zeta (5) \tau_2^{5/2}$ for a specific choice of $\mu_1$ that can be determined using the calculations in~\cite{Green:2008uj}.

For the genus 3 contribution, the analytic as well as non--analytic contributions
both have $\zeta (2)^3$ as their coefficient, hence a change of the scale simply shifts one contribution to another without changing the transcendentality. Thus the total contribution to $f^{(12)}_1$ from the genus 3 term that arises from the last line of \C{ans6loop} can be absorbed by redefining the scale of the logarithm and the overall coefficient of the $\zeta (2)^3 \tau_2^{-3/2}$ genus 3 analytic term. Exactly what fraction is absorbed by the finite piece and what fraction by the divergent piece does not follow from our analysis. This can be determined uniquely by calculating the string amplitude, which fixes unambiguously the infra--red finite and infra--red divergent parts, along with the scale $\mu_3$ of the logarithm.
So the last two terms in \C{ans6loop} can be ignored for our analysis.

Note that for genus greater than 0, the string amplitude always has infra--red divergences at sufficiently high orders in the momentum expansion, due to the contributions of the massless modes that propagate in the loop. Thus the amplitude splits into a sum of analytic and non--analytic parts. In string theory, this distinction is made unambiguously, and the non--analytic part of the amplitude is obtained from the degeneration limits of the moduli space of the various Riemann surfaces.  
While the momentum dependence of the analytic part is universal, that of the non--analytic part with its branch cut singularities, is dimension dependent. The branch cut singularity, for example, is logarithmic only in 10 dimensions for the terms we are discussing (generic terms in other dimensions have square root singularities). While the branch cut structure is determined directly by computing the string amplitude, it can also be determined by compactifying to lower dimensions, and looking at the terms that diverge on decompactification. These terms sum up to give the singularity structure of the amplitudes. So there is an unambiguous distinction between the analytic and non--analytic parts of the amplitude. For the logarithmic case relevant to our analysis, the distinction is fixed in string theory by calculating the scale, which is arbitrary in supergravity. The uniqueness of the scale follows from the modular invariance of the various multi--loop string amplitudes.     

Thus it is crucial that even though in 10 dimensions a change of the scale of the logarithm changes the coefficients of the analytic and non--analytic parts of the amplitude, there is no ambiguity in obtaining the coefficient in M theory, for the reasons described above. For example, for the genus one $s^6 \mathcal{R}^4$ interaction, the analytic part of the 10 dimensional amplitude is given by~\cite{Green:2008uj}\footnote{Here 
\be \s_2 = \Big(\frac{l_s}{2}\Big)^4 (s^2 + t^2 + u^2),\quad \s_3 = \Big(\frac{l_s}{2}\Big)^6 (s^3 + t^3 + u^3).\ee}     
\be \label{10d}I_{10}=  \frac{\pi}{3}\Big( \s_2^3 \frac{\zeta (3)^2}{30} + \s_3^2 \frac{61\zeta(3)^2}{1080}\Big).\ee 
When compactified on a circle of radius $r$, the part of the 9 dimensional amplitude which is linear in $r$ is 
\be I_9 = 2\pi r\cdot \frac{\pi}{3} \Big( \s_2^3 \frac{\zeta (3)^2}{30} + \s_3^2 \frac{61\zeta(3)^2}{1080}\Big)\ee
which precisely reproduces \C{10d} on decompactification. The other terms in the 9 dimensional amplitude either vanish on decompactification, or add to give the logarithmic terms in 10 dimensions. Hence, unlike in supergravity, there is no ambiguity.

Note that the infra--red divergent terms in \C{ans6loop} are given by
\be \eta_i \zeta (2) \Big( 2 \zeta (5) \tau_2^{5/2} + \frac{8}{3}\zeta (4) \tau_2^{-3/2} \Big){\rm ln}\tau_2 ,\ee 
which is non--perturbatively completed to $\eta_i E_{5/2} Y$ as demanded by unitarity and S--duality~\cite{Green:2006gt}.

Then the infra--red finite genus 1 and 3 contributions to $f_1^{(12)}$ are proportional to $\zeta (2) \zeta (3)^2$ and $\zeta (2)^3 (\Omega_1+\Omega_2\zeta (3))$ respectively for generic $\lambda_i$, where $\Omega_i$ has vanishing transcendentality.

While the perturbative contributions to the modular form $f_1^{(12)}$ take the form of \C{ans6loop} for generic choices of $\lambda_i$ in \C{eqn3}, one must separately solve \C{eqn3} when $\lambda_i= 63/4, 15/4$ and $-1/4$,
given the source terms in \C{manycont}. We now write down the solutions for these choices of $\lambda_i$. All the terms obtained from \C{ans6loop} after setting $c_{1,i} = c_{2,i} =0$ continue to exist in the solutions, and we do not mention them again for the sake of brevity. We only write down the new terms in the equations below.

\subsubsection{The analysis for $\lambda_i = 63/4$}

For $\lambda_i=63/4$, the solution to
\be \label{63/4eqn}4 \tau_2^2 \frac{\p^2}{\p \tau \p \bar\tau} h_1 = \frac{63}{4} h_1 -  \s_1 E_{3/2} f^{(6)} -\s_2 E_{3/2}^3 - \s_3 E_{5/2}Y\ee
is given by
\be \label{63/4}h_1 (\tau,\bar\tau) = c_0 \tau_2^{9/2} +c_4 \tau_2^{-7/2}-\frac{1}{2} \Big( \frac{\s_1}{3} + 2\s_2\Big) \zeta (3)^3 \tau_2^{9/2} {\rm ln} \tau_2 +\frac{16}{945}\zeta(2)^4 \s_1 \tau_2^{-7/2} {\rm ln}\tau_2 +\ldots,\ee
where the third term we have exhibited in \C{63/4} is a genus zero contribution that is inconsistent because of the logarithm. Hence
\be \s_1 = - 6\s_2.\ee
The last term in \C{63/4} is a genus 4 non--analytic contribution (which must be completed non--perturbatively) which must vanish. This is because the non--analytic contributions have to be of the form $E_{5/2} Y$ as demanded by unitarity, and so $\s_1 =0$.

Thus \C{63/4eqn} reduces to
\be \label{63/4main}4 \tau_2^2 \frac{\p^2}{\p \tau \p \bar\tau} h_1 = \frac{63}{4} h_1 - \s_3 E_{5/2}Y\ee

Note that the genus 3 contribution is $\sim \zeta(2)^3 (\Omega_1+\Omega_2 \zeta(3))$ as before\footnote{Of course, the  coefficients of the terms $\zeta(2)^3$ and $\zeta(2)^3 \zeta(3)$ depend on the explicit values of the various coefficients in the Poisson equations. We shall call them $\Omega_i$ as we do not determine their values.}. 

In the expression \C{eqn2} for $f_1^{(12)}$,  equation \C{63/4main} must be there in the sum as this is the only equation which can have a tree level contribution $\sim \zeta (9) \tau_2^{9/2}$ as demanded by \C{univ12}. All the other equations which do not have $63/4$ as the coefficient for $\lambda_i$, have a tree level contribution $\sim \zeta (3)^3 \tau_2^{9/2}$ from the source terms. Thus $c_0 = 2 \zeta (9)/27+ \psi \zeta (3)^3$, given \C{univ12}. Note that if $\s_3$ vanishes in \C{63/4}, then 
\be h_1 = \frac{1}{27}E_{9/2}= \frac{1}{27}\Big(2\zeta (9)\tau_2^{9/2} + 64\zeta (8) \tau_2^{-7/2}/35 +\ldots\Big), \ee 
as discussed in~\cite{Berkovits:1998ex}, and  thus $\psi =0$.
If $\s_3 \neq 0$, still we can obtain some constraint on $c_4$ using the analysis leading to \C{meqn}. For  $s=9/2$ in \C{meqn}, we get that
\be \label{c4}\zeta (9) \Big( -c_4 + \frac{64}{945}\zeta (8)\Big) +\frac{32}{35} \psi \zeta (3)^3 \zeta (8) = \ldots,\ee
where the $\ldots$ represent source term contributions $\sim \s_3$. Thus we see that $c_4 \sim \zeta (2)^4$, upto the terms we have calculated. This is of the same form as in \C{ans6loop}.

\subsubsection{The analysis for $\lambda_i = 15/4$}

For $\lambda_i=15/4$, the solution to
\be \label{15/4}4 \tau_2^2 \frac{\p^2}{\p \tau \p \bar\tau} h_2 = \frac{15}{4} h_2 -  \s_4 E_{3/2} f^{(6)} -\s_5 E_{3/2}^3 -\s_6 E_{5/2}Y\ee
is given by
\bea h_2 (\tau,\bar\tau) &=& c_1 \tau_2^{5/2} + c_3 \tau_2^{-3/2}  -4 \Big( \frac{\s_4}{3} + 3\s_5\Big) \zeta (2) \zeta (3)^2 \tau_2^{5/2}{\rm ln} \tau_2  \non \\ &&+  8 \Big( \frac{1}{5} \Big\{ 1+\frac{2}{189}\zeta (3) \Big\} \s_4 + 2\s_5\Big) \zeta(2)^3 \tau_2^{-3/2} {\rm ln} \tau_2 \non \\&&
-\frac{\s_6}{4} \zeta (2)\zeta (5) \tau_2^{5/2} ({\rm ln} \tau_2)^2  + \frac{\s_6}{3} \zeta (2) \zeta (4) \tau_2^{-3/2} {\rm ln} \tau_2+  \ldots.\eea
Thus $\s_6 =0$ because there is no $\tau_2^{5/2}({\rm ln} \tau_2)^2$ contribution to the genus 1 amplitude. Also because the genus 1 non--analytic amplitude is proportional to $\zeta (2) \zeta (5) \tau_2^{5/2} {\rm ln} \tau_2$~\cite{Green:2008uj}, we have that $\s_4 = -9\s_5$. Finally the genus 3 non--analytic amplitude proportional to $\tau_2^{-3/2} {\rm ln}\tau_2$ can have a lift to $E_{5/2} Y$ only if $\s_4=0$. Thus $\s_4 = \s_5 =0$, and \C{15/4} reduces to
\be 4\tau_2^2 \frac{\p^2}{\p \tau \p \bar\tau} h_2 = \frac{15}{4} h_2 ,\ee
which has the solution $h_2 \sim \zeta (2)E_{5/2} \sim \zeta (2)\zeta (5) \tau_2^{5/2} + \zeta (2)\zeta (4) \tau_2^{-3/2}$, which involves terms of the type given in the last line of \C{ans6loop}. As discussed above, they will not play a role in our analysis.

\subsubsection{The analysis for $\lambda_i = -1/4$}

For $\lambda_i=-1/4$, the solution to
\be 4 \tau_2^2 \frac{\p^2}{\p \tau \p \bar\tau} h_3 = -\frac{1}{4} h_3 -  \s_7 E_{3/2} f^{(6)} -\s_8 E_{3/2}^3 -\s_9 E_{5/2}Y\ee
is given by
\be \label{1/4}h_3 (\tau,\bar\tau) = (c_2 + \tilde{c}_2 {\rm ln} \tau_2)\tau_2^{1/2} -16 \Big(\frac{4\s_7}{15} + 3\s_8\Big) \zeta (2)^2 \zeta (3) \tau_2^{1/2} ({\rm ln}\tau_2)^2+  \ldots.\ee
 The non--analytic genus 2 contribution of the type $\tau_2^{1/2}{\rm ln}\tau_2$ must vanish as it does not have a lift to $E_{5/2} Y$, and so $\tilde{c}_2 = 0$ by unitarity. 

The other non--analytic genus 2 contribution which is of the form  $\zeta (2)^2 \zeta (3) \tau_2^{1/2} ({\rm ln}\tau_2)^2$ also must vanish due to constraints imposed by the 2 particle unitarity cut analysis, which yields all the logarithmic dependence. To see this, note that the discontinuity equation is given by (written very schematically)~\cite{Green:2006gt}
\bea \label{disc}
{\rm Disc}_s A^{\rm nonan}(p_i) \sim \int d^{10}k A^{(4)} (p_1, p_2, k, -k -p_1-p_2)A^{(4)} (-k,k-p_3-p_4,p_3,p_4) \non \\ \times\delta (k^2) \theta (k^0)\delta((k+p_1+p_2)^2) \theta((k+p_1+p_2)^0)\eea
which gives the $s$--channel discontinuity of the non--analytic part of the amplitude $A^{\rm nonan}$, where $A^{(4)}$ is the  on--shell 4--point amplitude. Inserting the various terms in the momentum expansion of the various string loop amplitudes to $A^{(4)}$, calculating \C{disc} and S--dualizing  gives the various non--analytic terms involving ${\rm ln}\tau_2$ terms we have discussed upto this order in the momentum expansion. 

To get terms $\sim ({\rm ln} \tau_2)^2$ from \C{disc}, we insert one factor of the genus one contribution to $A^{(4)}$ in \C{disc}. The leading contribution in the momentum expansion involves substituting the supergravity expression $A^{(4)}\sim \mathcal{R}^4/stu$ in the other vertex\footnote{For the genus one vertex, we insert $A^{(4)} \sim \zeta(2) \mathcal{R}^4$.}, leading to a non--analytic genus two term $\sim \zeta (2)^2 ({\rm ln} \tau_2)^2 s^5 \mathcal{R}^4$ term in the Einstein frame\footnote{This also follows from \C{f102}.}. The next contribution involves substituting the $\mathcal{R}^4$ vertex rather than the supergravity vertex leading to the non--analytic genus two term $\sim \tau_2^{3/2} \zeta (2)^2 \zeta (3) ({\rm ln} \tau_2)^2 s^8 \mathcal{R}^4$ in the Einstein frame\footnote{This also follows from \C{f16} from the source term $E_{3/2} Y^2$.} (another contribution comes from keeping the supergravity vertex, and using the factor $A^{(4)} \sim \zeta (3)\zeta (2) s^3 \mathcal{R}^4$ for the genus one vertex which follows from the expression for $f^{(6)}$ in \C{f6}). Thus a genus two term of the form $\zeta (2)^2 \zeta (3) \tau_2^{1/2} ({\rm ln} \tau_2)^2 s^6 \mathcal{R}^4$ in \C{1/4} is not obtained from the 2 particle unitarity cut analysis, and hence must vanish.

Hence
\be 4\s_7 + 45 \s_8 =0 \ee 
in \C{1/4}. Interestingly, then the analytic source term $\sim E_{3/2} f^{(6)} - 4E_{3/2}^3/45 \sim 0 \cdot \tau_2^{1/2} +\ldots$ which follows from \C{manycont}. Then $c_2$ in \C{1/4} is not constrained using \C{meqn}.

Note that the genus 3 contribution in \C{1/4} remains as before.
Thus it follows from our analysis that the analytic part of the genus 3 contribution to $f_1^{(12)}$ is proportional to 
\be \label{genus3} \zeta(2)^3 \Big( \Omega_1+\Omega_2 \zeta (3)\Big)\ee
where $\Omega_i$ has vanishing transcendentality. 

The analysis for $f_2^{(12)}$ in \C{nonuniv12} follows along similar lines, as the starting equation is \C{eqn3} with $f_{1,i}^{(12)} \rightarrow f_{2,i}^{(12)}$. 
Since the tree level contribution $\sim \zeta (9) \tau_2^{9/2}$ does not follow from the source terms, $f_2^{(12)}$ must have \C{63/4main} in the analogous decomposition \C{eqn2}.
The total tree level contribution from the source terms $E_{3/2} f^{(6)}$, $E_{3/2}^3$ and the part of $c_0$ in \C{63/4} that $\sim \zeta (3)^3$, to $f_2^{(12)}$ has to vanish as demanded by \C{nonuniv12}. 

Thus from \C{actM} and \C{actIIA}, based on our assumption we see that the M theory effective action has a term 
\be \label{act2}l_{11}^9 \zeta (2)^3\Big( \Omega_1 + \Omega_2 \zeta (3) \Big)\int d^{11}x \sqrt{-G} D^{12} \mathcal{R}^4  ,\ee
which is completely determined by the genus 3 contribution to the $D^{12} \mathcal{R}^4$ type II interaction. Now \C{act2} is true for both the $(s^3 + t^3 + u^3)^2 \mathcal{R}^4$ and $(s^2 + t^2 + u^2)^3\mathcal{R}^4$ interactions. Of course, the value of $\Omega_i$ need not be the same for both.  

\subsection{Non--BPS source terms}

So far we have analyzed terms in the type IIB effective action $S^{(n)}$ for $n \leq 9$. From the structure of the Poisson equations  we see that for these interactions the source terms are all BPS interactions (involving $S^{(n)}$ for $n \leq 6$), though the interactions in $S^{(n)}$ for $n \geq 7$ are non--BPS themselves. This structure changes immediately when we consider terms in the effective action $S^{(n)}$ for $n \geq 10$. The simplest example is provided by the coefficient function of the $D^{14} \mathcal{R}^4$ interaction which is in $S^{(10)}$, where the Poisson equation has the source term $E_{3/2} f^{(8)}$ (see \C{f14} for details) where $f^{(8)}$ is the $D^8 \mathcal{R}^4$ coupling which is in $S^{(7)}$.  
This leads to a more complicated expression for the various coefficient functions because the source terms themselves are non--BPS. 

In fact, there is yet another source of complication that arises. In $S^{(n)}$ for $n \geq 7$, there are new interactions with new $SL(2,\mathbb{Z})$ invariant coefficient functions that can contribute as source terms. These non--BPS interactions have hardly been studied, and we briefly mention their origin for the known cases. The simplest example of such an interaction which gives genus 0 contributions that is different from those in \C{tree} is provided by $f^{(14;5)}$, the coefficient function of the $D^{14} \mathcal{R}^5$ interaction which is in $S^{(11)}$ (see \C{new} for details)~\cite{Schlotterer:2012ny}. Given the structure of \C{inv} and \C{sum}, the $f^{(14;5)}$ interaction can yield $E_{3/2} f^{(14;5)}$ source terms in the Poisson equations for the interactions in $S^{(14)}$. 

The simplest example of such a coefficient function $f^{(8;5)}$ in 10 dimensions which vanishes at genus 0 but has a non--vanishing genus 1 analytic contribution is the coupling of the $D^8 \mathcal{R}^5$ interaction which is in $S^{(8)}$~\cite{Green:2013bza} (see the discussion below \C{f16}), and can yield $E_{3/2} f^{(8;5)}$ source terms for $S^{(11)}$~\footnote{The genus zero and the analytic part of the genus one $D^6 \mathcal{R}^5$ interaction in $S^{(7)}$ vanishes in 10 dimensions, and hence its leading analytic contribution is at genus two, and so the analytic part of its coefficient function $f^{(6;5)} \sim \tau_2^{-1/2}$ at weak coupling.}.  Clearly this structure is very general and the structure of the source terms gets complicated as one increases $n$ because of the added contribution of many new non--BPS couplings.        

Thus our analysis of considering the source term contributions to constrain various string loop amplitudes does not capture all the terms due to lack of information about the various source terms. Hence this does not give us the complete answer for the terms of various transcendentality for the M theory interactions. However, the various source terms we can analyze still provide a lot of information about various string loop amplitudes in the type IIB theory, as well as M theory interactions. So for illustrative purposes, for the source terms we can analyze we schematically describe below some details of the $D^{18} \mathcal{R}^4$ and $D^{24} \mathcal{R}^4$ interactions in the type IIB theory, as well as M theory. This gives us non--trivial information about string amplitudes at various high genera in the type IIB  theory which are quite difficult to obtain in general.    
It would be quite challenging to understand the structure of these source terms in detail, at least for moderate values of $n$.     

Thus in the discussion below, our aim is not to obtain the complete amplitude, but to use constraints of supersymmetry and S--duality to determine some of the terms that contribute to the amplitude, given the various source terms we can analyze. 

\subsection{Schematics of the $D^{18} \mathcal{R}^4$ interaction}

We now consider the case for $n=3$, which is the $D^{18} \mathcal{R}^4$ interaction. We shall be schematic and brief in our discussion. Also given the Poisson equations with several undetermined coefficients we can solve it, as in the analysis for the $D^{12} \mathcal{R}^4$ interaction (\C{ans6loop} for example), to obtain various string amplitudes at various genera upto overall factors, which are both analytic as well as non--analytic in the complex modulus $\tau$. For brevity, we shall simply write down the various coefficients of the various higher genus amplitudes that we obtain. Our primary focus will be to obtain the contribution to the genus 4 amplitude.

For the $D^{18} \mathcal{R}^4$ interaction which is in $S^{(12)}$, from \C{inv} and \C{sum} we have that
\be \delta^{(0)} S^{(12)} + \delta^{(12)} S^{(0)} + \delta^{(3)} S^{(9)} +\delta^{(9)} S^{(3)} +\delta^{(4)} S^{(8)} + \delta^{(8)} S^{(4)} + \delta^{(5)} S^{(7)} + \delta^{(7)} S^{(5)} +\delta^{(6)}S^{(6)}=0.\ee

The source terms involve both $S^{(7)}$ and $S^{(8)}$, which include the $D^8 \mathcal{R}^4$ and $D^{10} \mathcal{R}^4$ interactions respectively, which we briefly discuss in appendix B.

Now for the $D^{18} \mathcal{R}^4$ term, let us first consider the universal interaction whose coefficient is the modular form $f_1^{(18)}$ given to leading order in the weak coupling expansion by \C{univ18}.
Thus writing
\be \label{18}f_1^{(18)} = \sum_i f_{1,i}^{(18)},\ee
we see that each $f_{1,i}^{(18)}$ satisfies the Poisson equation 
\bea \label{18more}4 \tau_2^2 \frac{\p^2}{\p \tau \p \bar\tau} f_{1,i}^{(18)} &=& \hat\lambda_i f_{1,i}^{(18)} + E_{3/2} \sum_{j=1}^2 \sum_k \alpha_{ijk}  f_{j,k}^{(12)} + \beta_i E_{3/2}^2 f^{(6)} + \gamma_i Y E_{3/2} E_{5/2} \non \\ &&+Y \sum_j \epsilon_{ij} f^{(10)}_j + \omega_i Y^3  + \eta_i \Big(f^{(6)} \Big)^2 + E_{5/2}\sum_j \kappa_{ij}  f^{(8)}_j . \eea

From \C{18more}, knowing the various perturbative contributions to the various modular forms involved in the source terms, one can immediately write down a plethora of  perturbative contributions to $f_{1,i}^{(18)}$ for generic values of $\hat\lambda_i$ consistent with string perturbation theory. The analytic and non--analytic contributions at various genera involve linear combinations of the following\footnote{There are additional contributions given in \C{newadd} from more amplitudes considered later.}:

\vspace{.2cm}

(i) Genus 0: \be \zeta (3)^4 \tau_2^6, \quad \zeta (3) \zeta (9) \tau_2^6, \quad \zeta (5) \zeta (7) \tau_2^6 \non \ee

(ii) Genus 1: \be  \zeta (2) \zeta (3) \zeta (5) \tau_2^4,  \quad \zeta (2) \zeta (3)^3 \tau_2^4, \quad \zeta (2) \zeta (9) \tau_2^4 , \quad \zeta (2) \zeta (3) \zeta (5) \tau_2^4 {\rm ln} \tau_2, \non \ee 

(iii) Genus 2: \be \zeta(2)^2 \zeta (5) \tau_2^2 ,\quad\zeta (2)^2\zeta (3)^2 \tau_2^2, \quad \zeta (2)^2 \zeta (7) \tau_2^2 , \quad \zeta (2)^2 \zeta (5) \tau_2^2 {\rm ln} \tau_2, \non \ee

(iv) Genus 3: \bea &&\zeta (2)^3 , \quad \zeta (2)^3 \zeta (3) , \quad \zeta (2)^3 \zeta (5), \quad \zeta (2)^3 \zeta (3)^2 \non \\ &&\zeta (2)^3 ({\rm ln} \tau_2)^3, \quad  \zeta (2)^3 ({\rm ln} \tau_2)^2, \quad \zeta (2)^3 {\rm ln} \tau_2, \quad \zeta (2)^3 \zeta (3) {\rm ln} \tau_2, \non\eea

(v) Genus 4: \be \zeta (2)^4 \tau_2^{-2}, \quad \zeta (2)^4 \zeta (3) \tau_2^{-2}, \quad \zeta (2)^4 \zeta (5) \tau_2^{-2}, \quad \zeta (2)^4 \tau_2^{-2} {\rm ln} \tau_2 \non \ee

(vi) Genus 5: \be \zeta (2)^5 \tau_2^{-4}\non \ee

(vii) Genus 6: \be\zeta (2)^6 \tau_2^{-6}, \quad \zeta (2)^6 \zeta (3) \tau_2^{-6}, \quad \zeta (2)^6\tau_2^{-6} {\rm ln}\tau_2 \non \ee

(viii) Genus 7: \be \zeta (2)^7 \tau_2^{-8} \non\ee

(ix) Genus 9: \be \label{hg} \zeta (2)^9 \tau_2^{-12} \ee

\vspace{.2cm}

In obtaining the above expressions, the known perturbative contributions deduced earlier has been used, and there are several terms that give the same contribution. In the list above, we have mentioned the genus 2 and 3 contributions $\zeta (2)^2 \zeta (5) \tau_2^2$ and $\zeta (2)^3\zeta (5)$ respectively which come from $E_{5/2} f^{(8)}_{i}$
and involve the genus 2 and 3 contributions $a^{(8)}_{2,i} \tau_2^{-1/2}$ and $a^{(8)}_{3,i}$ to $f^{(8)}_{i}$, using \C{needit}, \C{3/4} and \C{a83}\footnote{They also yield genus 4 and 5 contributions $\zeta(2)^4 \tau_2^{-2}$ and $\zeta (2)^5 \tau_2^{-4}$ respectively to \C{hg}.}.  Also at genus 4, there is a contribution coming from $Y f^{(10)}_i$ of the form $\zeta (2) a^{(10)}_{3,i}$ which gives $\zeta (2)^4$ using the genus 3 contribution to $f^{(10)}_i$  (see \C{f102}). For the genus 9 contribution, we have used \C{genusn}. Finally, there are some analytic contributions at various genera in \C{hg} which could possibly be absorbed in the logarithmic scale of some non--analytic terms in a modular invariant way in the total contribution, in which case those contributions vanish. We have kept these terms because of the lack of this information about higher genus amplitudes\footnote{Whether the $\zeta (2) \zeta (3) \zeta (5) s^9 \mathcal{R}^4$ analytic genus 1 contribution that follows from \C{hg} is absorbed by the scale of the logarithm of the non--analytic $\zeta (2) \zeta (3) \zeta (5) s^9 {\rm ln} (-\mu l_s^2 s) \mathcal{R}^4$ contribution should be verifiable by generalizing \cite{Green:2008uj} to higher orders in the momentum expansion.}.

As discussed before, there can be several other possible source terms in \C{18more}. For example, including the contributions from the $D^{2k} \mathcal{R}^5$ interactions, the possible source terms are of the form $E_{3/2} f^{(10;5)}$, $Y f^{(8;5)}$ and $E_{5/2} f^{(6;5)}$ involving $f^{(10;5)}, f^{(8;5)}$ and $f^{(6;5)}$, the coefficient functions of the $D^{10} \mathcal{R}^5, D^8 \mathcal{R}^5$ and $D^6 \mathcal{R}^5$ interactions respectively. Proceeding along similar lines, one can consider the $D^{2k} \mathcal{R}^6$ interactions and so on, to obtain more possible source terms. The exact role of these source terms in the various Poisson equations should follow from analyzing the maximally fermionic interactions in these supermultiplets along the lines of~\cite{Basu:2008cf}. 

Note that the various non--analytic terms in \C{18more} that lead to \C{hg} are of the form such that their total contribution can be non--perturbatively completed to give the non--local term in the type IIB effective action in the Einstein frame of the form
\be  Y\Big( f^{(10)} + E_{3/2} E_{5/2} + Y^2 \Big) s^9 \mathcal{R}^4,\ee  
 which follows from generalizing~\cite{Green:2006gt}. Hence the 2 particle unitarity cut analysis also implies that the total contribution of the form $Y (f^{(8;5)} +\ldots)s^9 \mathcal{R}^4$ must vanish, where the $\ldots$ stands for other contributions from $S^{(8)}$.

The analysis for $f_{2,i}^{(18)}$ proceeds along similar lines. On adding up all the contributions, the term $\sim \zeta (5) \zeta (7) \tau_2^6$ must vanish for $f_1^{(18)}$, while the term $\sim \zeta (3)^4 \tau_2^6$ must vanish for $f_2^{(18)}$, to be consistent with \C{univ18} and \C{nonuniv18}.

Keeping only the contributions in \C{hg}, our analysis suggests that the $D^{18} \mathcal{R}^4$ interaction in the M theory effective action has at least the terms
\be \label{act3}l_{11}^{15} \zeta (2)^4 \Big( \tilde\Omega_1 +\tilde\Omega_2 \zeta (3) +\tilde\Omega_3 \zeta (5)\Big) \int d^{11}x \sqrt {-G} D^{18} \mathcal{R}^4\ee
where the constants $\tilde\Omega_i$ have vanishing transcendentality. This is true individually for each distinct spacetime structure of the interaction. It would be very interesting to see how the other possible contributions to \C{act3} change the structure.

\subsection{Schematics of the $D^{24} \mathcal{R}^4$ interaction}

Finally, we shall very briefly outline the details for the $n=4$ case of the $D^{24} \mathcal{R}^4$ interaction. From \C{inv} and \C{sum}, we have that 
\bea \label{longe}&&\delta^{(0)} S^{(15)} + \delta^{(15)} S^{(0)} + \delta^{(3)} S^{(12)} + \delta^{(12)} S^{(3)} + \delta^{(4)} S^{(11)} + \delta^{(11)} S^{(4)} \non \\ &&+ \delta^{(5)} S^{(10)} + \delta^{(10)} S^{(5)} + \delta^{(6)} S^{(9)} + \delta^{(9)} S^{(6)} + \delta^{(7)} S^{(8)} + \delta^{(8)} S^{(7)}=0,\eea
where we shall restrict ourselves to source terms from the $D^{2k} \mathcal{R}^4$ interactions.
In \C{longe}, the source terms involve $S^{(11)}$ and $S^{(10)}$, which include the $D^{16} \mathcal{R}^4$ and $D^{14} \mathcal{R}^4$ interactions respectively, which we briefly discuss in appendix B.

Now from \C{24form}, we see that each $f_i^{(24)}$ ($i=1,2,3$) must split into a sum of the form
\be f_i^{(24)} = \sum_{j} f_{i,j}^{(24)},\ee
where each $f_{i,j}^{(24)}$ satisfies the Poisson equation. For example, $f_{1,i}^{(24)}$ satisfies the Poisson equation 
\bea \label{f24}4\tau_2^2 \frac{\p^2}{\p \tau \p \bar\tau} f_{1,i}^{(24)} = \lambda_i f_{1,i}^{(24)} + E_{3/2} \sum_{j=1}^2 \sum_k \tilde\alpha_{ijk}  f^{(18)}_{j,k} +E_{3/2}^2 \sum_{j=1}^2 \sum_k \tilde\beta_{ijk} f^{(12)}_{j,k} + Y E_{3/2}\sum_j \tilde\gamma_{ij} f^{(10)}_j \non \\ + E_{3/2} \Big( f^{(6)} \Big)^2 + E_{3/2} E_{5/2} \sum_j \tilde\epsilon_{ij} f^{(8)}_j + Y \sum_{j=1}^2 \sum_k \tilde\eta_{ijk} f^{(16)}_{j,k} + Y^2 \sum_j \tilde\lambda_{ij} f^{(8)}_{j} + \theta_i E_{5/2}  Y f^{(6)}\non \\ + E_{5/2} \sum_j \omega_{ij} f^{(14)}_{j} +\xi_i E_{5/2}^3  + f^{(6)} \sum_{j=1}^2 \sum_k \psi_{ijk} f^{(12)}_{j,k} + \sum_{j,k} \Upsilon_{ijk} f^{(8)}_j f^{(10)}_k . \non \\\eea

The genus zero contributions with coefficients $\zeta(3)^5, \zeta(3)^2 \zeta (9), \zeta (5)^3$ and $\zeta (3) \zeta (5) \zeta (7)$ follow immediately from \C{f24}, in agreement with $f^{(24)}_1$ and $f^{(24)}_3$, while the term involving $\zeta (15)$ is not obtained from the source terms in \C{f24}. This must arise from \C{f24} with $\lambda_i = 195/4$, and similarly for the equations for $f^{(24)}_2$ and $f^{(24)}_3$.

As in the earlier cases, one can immediately write down several amplitudes at various genera which are analytic as well as non--analytic in the complex coupling $\tau$. However, for the sake of brevity, we shall only write down the analytic part of the genus 5 amplitude which is relevant for the nature of the M theory interaction. The various terms involved have different transcendentalities, and we write down terms only upto transcendentality 19. One can generalize to terms having higher transcendentality by considering more terms at various genera for the non--BPS couplings. 

Based on the various known amplitudes we have already analyzed before and their contributions to the source terms in \C{f24}, this is given by 
\be \label{genus5}\zeta (2)^5 \Big( 1 +\zeta (3) +\zeta (5) +\zeta (7)\Big) \tau_2^{-5/2} +\zeta (3)^2 a^{(12)}_{5} \tau_2^{-5/2} +\zeta (3) \zeta (5) a^{(8)}_5 \tau_2^{-5/2},\ee
where $a^{(12)}_5$ and $a^{(8)}_5$ are the coefficients of the genus 5 amplitudes for the $D^{12} \mathcal{R}^4$ and $D^8 \mathcal{R}^4$ interactions respectively (i.e., $f^{(12)}_{i,j} \sim a^{(12)}_5 \tau_2^{-11/2}$ and $f^8_i \sim a^{(8)}_5 \tau_2^{-13/2}$). Keeping only terms that are relevant to have terms upto transcendentality 19 in the final answer, we show in appendix C that $a^{(8)}_5 \sim \zeta (2)^5$ and $a^{(12)}_5 \sim \zeta (2)^5 + \zeta (2)^5 \zeta (3)$ (see \C{g58} and \C{g512}), and thus \C{genus5} gives us
\be \label{ge5}
\zeta (2)^5\Big( \underline\Omega_1 +\underline\Omega_2 \zeta (3) +\underline\Omega_3\zeta (5) +\underline\Omega_4\zeta(3)^2+\underline\Omega_5\zeta (7) +\underline\Omega_6\zeta (3)\zeta (5) +\underline\Omega_7 \zeta (3)^3  \Big)\tau_2^{-5/2}.\ee
Thus our analysis suggests that the $D^{24} \mathcal{R}^4$ interaction in the M theory effective action has at least the terms
\be \label{act4}l_{11}^{21} \zeta(2)^5 \Big( \underline\Omega_1 +\underline\Omega_2 \zeta (3) +\underline\Omega_3\zeta (5) +\underline\Omega_4\zeta(3)^2+\underline\Omega_5\zeta (7) +\underline\Omega_6\zeta (3)\zeta (5) +\underline\Omega_7 \zeta (3)^3\Big) \int d^{11}x \sqrt{-G} D^{24} \mathcal{R}^4,\ee
for each independent spacetime structure in the $D^{24}\mathcal{R}^4$ interaction, where $\underline\Omega_i$ has vanishing transcendentality.

Though we have restricted ourselves to  only a certain subset of source terms in the Poisson equation, given the genus one contribution to $f^{(8;5)}$ (see the discussion below \C{f16}) it follows that we can have a source term $f^{(8)} f^{(8;5)}$ in \C{f24} which produces a genus 5 contribution $\sim \zeta (2)^4 \tau_2^{-9/2} \times a^{(8;5)}_1\tau_2^2 \sim \zeta (2)^4 a^{(8;5)}_1 \tau_2^{-5/2}$, where the genus 1 coefficient $a^{(8;5)}_1 \sim \zeta (2) \zeta (5) +\Upsilon$ on using \C{extra}. Thus we see very easily how new non--BPS interactions enter into the analysis.

\subsection{Some generalities}

For the $D^{30} \mathcal{R}^4$ interaction in M theory, it is easy to see that it is possible to have a MZV in the coefficient. Since the type IIB $D^{30} \mathcal{R}^4$ interaction is in $S^{(18)}$, the Poisson equation has $f^{(16)} f^{(8)}$ as source terms. This yields a genus 6 contribution of the form 
\be \label{t11}f^{(16)} f^{(8)} \sim \Big(\zeta (11) + \zeta (3)^2 \zeta (5)\Big) \tau_2^{11/2} \times \zeta (2)^6 \tau_2^{-17/2} \sim \zeta (2)^6\Big(\zeta (11) + \zeta (3)^2 \zeta (5)\Big) \tau_2^{-3}\ee
on using \C{a86}. However, one can have a new kind of contribution in addition to \C{t11} which comes from the $f^{(14;5)} f^{(8)}$ source term in the Poisson equation, where $f^{(14;5)}$ is the coefficient function of the $D^{14} \mathcal{R}^5$ interaction. Now apart from $\zeta (11)\tau_2^{11/2}$ and $\zeta(3)^2 \zeta (5)\tau_2^{11/2}$ at genus 0, $f^{(14;5)}$ has additional terms given by~\cite{Schlotterer:2012ny} 
\be \label{new}
T(11) \tau_2^{11/2}\equiv \Big( 9\zeta (2) \zeta (9)+\frac{6}{25}\zeta (2)^2 \zeta (7)-\frac{4}{35}\zeta(2)^3 \zeta (5)+\frac{1}{5} \zeta(3,3,5) \Big) \tau_2^{11/2}\ee 
where $\zeta (3,3,5)$ is the Multiple Zeta Value (MZV) of depth 3 defined by
\be \zeta (3,3,5) = \sum_{0<m<n<p} \frac{1}{m^3 n^3 p^5}.\ee
Apart from the MZV, the $D^{14} \mathcal{R}^5$ is the simplest multi--graviton amplitude where zeta functions of even transcendentality arise as genus 0.   
Thus in addition to \C{t11} we can
also have a contribution of the form
\be \label{addnew} T(11) \zeta (2)^6 \tau_2^{-3}, \ee
 where $T (11)$ is given by \C{new}. If the total contribution of the type \C{addnew} is non--vanishing, it contributes to the coefficient of the $D^{30} \mathcal{R}^4$ term in the M theory effective action, along with \C{t11}.

Though we do not expect the strong coupling expansion of the type IIA amplitude to be qualitatively related to the weak coupling expansion for arbitrary values of $n$, it is worthwhile to discuss the general structure of of the M theory interactions keeping $n$ arbitrary, to see what we get when the weak coupling expansion does qualitatively reproduce the strong coupling result, in the sense of transcendentality.   
As in our analysis above, at each step we have to use the expressions for the various couplings we have obtained in the previous step. Thus recursively using the structure of the Poisson equations, one can constrain higher genus string amplitudes, and consequently the nature of the coefficient of the M theory interaction. However, the lack of understanding of the various non--BPS source terms in the Poisson equation makes the analysis rather difficult as $n$ increases. Even then, our analysis shows that there is a systematic analysis which leads to a structured set of equations for determining them, which is a consequence of supersymmetry and S--duality. Thus our analysis suggests that for small values of $n$, the $D^{6n} \mathcal{R}^4$ term in the M theory effective action is of the  form  
\be \label{genn} l_{11}^{6n-3} \zeta (2)^{n+1} \Big( \Sigma_1 + \S_2 \zeta (3) +\S_3 \zeta (5) +\ldots\Big)\int d^{11}x \sqrt{-G} D^{6n} \mathcal{R}^4, \ee
 where the constants $\Sigma_i$ have vanishing transcendentality. We have only considered the $t_8 t_8 R^4$ part of the $\mathcal{R}^4$ interaction for simplicity. However, the structure of the transcendentality of the coefficients obtained for the various interactions in M theory which follows naturally suggests that the structure of \C{genn} might also be true for the $\epsilon_{10} \epsilon_{10} R^4$ part of the $\mathcal{R}^4$ interaction for small values of $n$.

We have analyzed only a subset of the local purely gravitational interactions in the effective action. This is only a very small subset of all the possible interactions, and it would be interesting to generalize the analysis to interactions involving the 4--form $G_4$ of 11 dimensional supergravity as well, along the lines of~\cite{Hyakutake:2007vc,Liu:2013dna} where some of the interactions at the 8 derivative level have been constructed.

\vspace{.2cm}

{\bf{Acknowledgements:}} I am thankful to Michael Green and Pierre Vanhove for useful comments.

\section{Appendix}

\appendix

\section{Transcendentality from integration by parts}

In the main text, we sometimes require to know the coefficient of certain high genus amplitudes (at least the transcendentality) for the various interactions in the type IIB theory which are not known directly from perturbative string calculations. Also given the limitations of our understanding of the various modular forms, we do not know the complete answer. However, based on the structure of the Poisson equations, we can impose some constraint on the high genus amplitude. We describe the method below, which is a generalization of~\cite{Green:2005ba}. 

Let the Poisson equation which the $SL(2,\mathbb{Z})$ modular form $f(\tau,\bar\tau)$ satisfies be
\be \label{integrate} 4 \tau_2^2 \frac{\p^2}{\p \tau \p \bar\tau} f = s(s-1) f + \mathcal{S},\ee 
where $\mathcal{S}$ is the total contribution from the various source terms. If $\mathcal{S}=0$, then we have that
\be f(\tau,\bar\tau) = E_s (\tau,\bar\tau) = 2\zeta (2s) \tau_2^s + 2\sqrt{\pi} \frac{\Gamma(s-1/2)\zeta (2s-1)}{\Gamma(s)}\tau_2^{1-s}+\ldots,\ee
which satisfies 
\be \label{Es}4\tau_2^2 \frac{\p^2}{\p \tau \p \bar\tau}E_s= s(s-1) E_s.\ee
For the cases we need, the source terms might or might not vanish, and the explicit expressions for some of the source terms are not known. Still we can obtain some constraint on the perturbative structure of $f$. To see this, we convolute \C{integrate} with $E_s$ and integrate over $\mathcal{F}$, the fundamental domain of $SL(2,\mathbb{Z})$, to get that
\be \label{int1}\int_{\mathcal{F}} \frac{d^2\tau}{\tau_2^2} E_s \Big(4\tau_2^2\frac{\p^2}{\p \tau \p \bar\tau}f \Big) = s(s-1) \int_{\mathcal{F}} \frac{d^2 \tau}{\tau_2^2} E_s f + \int_{\mathcal{F}} \frac{d^2 \tau}{\tau_2^2} E_s \mathcal{S} . \ee 
Integrating by parts, using \C{Es} and recalling that the only boundary of $\mathcal{F}$ is at $\tau_2 \rightarrow \infty$, we get that
\be \label{int2} \Big( E_s \frac{\p f}{\p \tau_2} - \frac{\p E_s}{\p \tau_2} f\Big)\Big\vert_{\tau_2 \rightarrow \infty} = \int_{\mathcal{F}} \frac{d^2 \tau}{\tau_2^2} E_s \mathcal{S}.\ee
Note that on the left hand side of \C{int2} the non--perturbative contributions which $\sim e^{-2\pi \vert k \vert \tau_2} \rightarrow 0$ (the fall--off is faster for multi--instanton bound states), and so only the perturbative contributions survive. In fact, substituting the various such contributions leads to terms which are finite, as well as terms which diverge as $\tau_2 \rightarrow \infty$. To keep track of the exact nature of the various terms, we simply replace $\mathcal{F}$ by $\mathcal{F}_L$ in \C{int1}, where $\mathcal{F}_L$ is the truncated fundamental domain of $SL(2,\mathbb{Z})$ with $\tau_2 \leq L$. Then \C{int2} reduces to
\be \label{mainint} \Big( E_s \frac{\p f}{\p \tau_2} - \frac{\p E_s}{\p \tau_2} f\Big)\Big\vert_{\tau_2 = L \rightarrow \infty} = \int_{\mathcal{F}_L} \frac{d^2 \tau}{\tau_2^2} E_s \mathcal{S}.\ee 
We can now evaluate both sides of \C{mainint} and equate the terms with distinct $L$ dependences\footnote{The right hand side of \C{mainint} can only be evaluated in simple cases, on using the Rankin--Selberg unfolding technique. For example, this was used to obtain the genus 3 amplitude for the $D^6 \mathcal{R}^4$ interaction in~\cite{Green:2005ba}.}.      

Let us consider the terms which are independent of $L$ on the left hand side of \C{mainint}. Among the various perturbative contributions to $f$, they only involve
\be f = a_1 \tau_2^s + a_2 \tau_2^{1-s}+\ldots,\ee 
leading to
\be \label{meqn}2(1-2s) \zeta (2s) a_2 + 4\sqrt{\pi} \frac{\Gamma(s+1/2) \zeta (2s-1)}{\Gamma(s)}a_1 = \int_{\mathcal{F}_L} \frac{d^2 \tau}{\tau_2^2} E_s \mathcal{S}\Big\vert_{\rm finite}.\ee
We consider several cases where we have some information about the transcendentality of some of the contributions to $a_1$ and $a_2$, which constrains the transcendentality of the unknown contributions.  

\section{Schematics of the $D^8 \mathcal{R}^4, D^{10} \mathcal{R}^4, D^{14} \mathcal{R}^4$ and $D^{16} \mathcal{R}^4$ interactions}

\subsection{The $D^8 \mathcal{R}^4$ and $D^{10} \mathcal{R}^4$ interactions}

The $D^8 \mathcal{R}^4$ and $D^{10} \mathcal{R}^4$ couplings appear as source terms in the Poisson equations for the $D^{18} \mathcal{R}^4$ coupling, and thus we briefly discuss their structure based on \C{inv} and \C{sum}.

Let $f^{(8)} (\tau,\bar\tau)$ and $f^{(10)}(\tau,\bar\tau)$ be the moduli dependent couplings of the $D^8 \mathcal{R}^4$ and $D^{10} \mathcal{R}^4$ interactions respectively, which come with the unique spacetime structures $(s^2 + t^2 + u^2)^2 \mathcal{R}^4$ and $(s^2 + t^2 + u^2) (s^3 + t^3 + u^3)\mathcal{R}^4$ respectively. We can expand them as 
\bea \label{2sum} f^{(8)} = \sum_i f^{(8)}_i, \non \\ f^{(10)} =\sum_i f^{(10)}_i,\eea
where each $f^{(8)}_i$ satisfies
\be \label{f8} 4 \tau_2^2 \frac{\p^2}{\p \tau \p \bar\tau} f^{(8)}_i = \kappa_i f^{(8)}_i + \epsilon_i E_{3/2} Y,\ee
while each $f^{(10)}_i$ satisfies
\be \label{f10} 4 \tau_2^2 \frac{\p^2}{\p \tau \p \bar\tau} f^{(10)}_i =\omega_i f^{(10)}_i + \xi_i E_{3/2}E_{5/2} + \psi_i Y^2,\ee
where $\kappa_i,\epsilon_i,\omega_i,\xi_i$ and $\psi_i$ are numbers of vanishing transcendentality. Now \C{f8} and \C{f10} follow from
\be \delta ^{(0)} S^{(7)} +\delta^{(7)} S^{(0)} + \delta^{(3)} S^{(4)} +\delta^{(4)} S^{(3)} =0,\ee
and
\be \delta^{(0)} S^{(8)} +\delta^{(8)} S^{(0)} +\delta^{(3)} S^{(5)} +\delta^{(5)} S^{(3)} +\delta^{(4)} S^{(4)} =0,\ee
respectively. As a very simple consistency check, from \C{f8} note that for $f^{(8)}_i$ for generic $\kappa_i$, this implies that
\be \label{needit} f^{(8)}_i \sim \epsilon_i \zeta (2) \Big(\zeta (3) \tau_2^{3/2} +\zeta (2) \tau_2^{-1/2}\Big){\rm ln}\tau_2 + \epsilon_i \zeta (2)\Big(\zeta (3) \tau_2^{3/2} + \zeta (2) \tau_2^{-1/2}\Big),\ee 
which are genus 1 and 2 analytic as well as non--analytic terms. The non--analytic terms are consistent with unitarity argument~\cite{Green:2006gt}, as they get non-perturbatively completed to $E_{3/2} Y$. In fact the non--analytic genus 1 amplitude has precisely this structure, while the total contribution of the analytic part of the genus 1 amplitude must vanish~\cite{Green:2008uj}\footnote{This holds for the logarithmic scale defined by
\be {\rm ln}~ \mu_1 = \frac{9}{10} -{\rm ln}~\Big(\frac{2}{\pi e^{-\gamma}}\Big) + \frac{\zeta '(3)}{\zeta (3)} -\frac{\zeta '(4)}{\zeta (4)}\ee 
along the lines of the discussion following \C{ans6loop}.}. The analytic part of the genus 2 amplitude is thus $\sim \zeta (2)^2\tau_2^{-1/2}$.

Of course, the total contribution to the genus 1 and 2 amplitudes involves the ones given by \C{needit} for generic $\kappa_i$ where they arise only from the source terms, as well as the ones obtained from solving \C{f8} with $\kappa_i=3/4$, where they arise also from the homogeneous part of the Poisson equation. For $\kappa=3/4$, let the equation be
\be \label{3/4} 4\tau_2^2 \frac{\p^2}{\p \tau \p \bar\tau}q = \frac{3}{4} q + \mu E_{3/2} Y.\ee  
Thus, the perturbative part of $q$ is given by
\bea \label{vanish} q (\tau,\bar\tau)&=& c_1 \tau_2^{3/2} + c_2 \tau_2^{-1/2} -\frac{1}{2} \zeta (2) \zeta (3) \mu \tau_2^{3/2} {\rm ln} \tau_2 -\zeta(2)^2 \mu \tau_2^{-1/2} {\rm ln}\tau_2 \non \\ &&+\frac{1}{2} \zeta (2) \zeta (3)\mu \tau_2^{3/2} ({\rm ln}\tau_2)^2 - \zeta(2)^2 \mu \tau_2^{-1/2} ({\rm ln}\tau_2)^2. \eea   
The last two terms in \C{vanish} are inconsistent with unitarity because the leading $({\rm ln} s)^2$ term is $({\rm ln} s)^2s^5 \mathcal{R}^4$, hence $\mu =0$.
Hence 
\be q \sim \zeta (2) E_{3/2} = 2\zeta (2) \zeta (3) \tau_2^{3/2} + 4\zeta (2)^2 \tau_2^{-1/2}+\ldots , \ee
 which yields analytic genus 1 and 2 contributions as given in \C{needit}.
Thus, if the analytic part of the genus 2 contribution is non--vanishing, it is $\sim \zeta (2)^2 \tau_2^{-1/2}$. It would be interesting to calculate the string amplitude to see if the choice of the logarithmic scale determined by string theory forces this coefficient to vanish or not.

It is also be very useful for our purposes that the genus 4 contribution to $f^{(8)}_i \sim \zeta (8) \tau_2^{-9/2}$, as we are considering the $D^8 t_8 t_8 R^4$ interaction\footnote{From \C{genusn}, the genus 4 type IIA $D^{8} t_8t_8R^4$ amplitude is equal to
\bea \frac{20}{8} \zeta (8) e^{6\phi_A} l_s^{8} \mathcal{W}^4 \mathcal{R}^4,\eea
while the genus 5 type IIA $D^{10} t_8t_8R^4$ amplitude is equal to
\bea \label{g5} \frac{99}{25} \zeta (10) e^{8\phi_A} l_s^{10} \mathcal{W}^5 \mathcal{R}^4,\eea
where we have used 
\be \zeta (6) = \frac{\pi^6}{945}, \quad \zeta (8)=\frac{\pi^8}{9450}, \quad \zeta (10)=\frac{\pi^{10}}{93555}.\ee}.

In the main text, we shall also find the expression for the genus 3  amplitude to be useful, hence we analyze it. 
From \C{f8}, we see that the genus 3 amplitude must have $\kappa_i = 35/4$, and so must be contained in the equation  
\be \label{g38}4\tau_2^2 \frac{\p^2}{\p \tau \p \bar\tau} \tilde{q} = \frac{35}{4} \tilde{q} +\tilde\mu E_{3/2} Y.\ee 
If $\tilde\mu =0$, \C{g38} has the solution~\cite{Berkovits:1998ex}
\be \tilde{q} = E_{7/2} = 2\zeta (7) \tau_2^{7/2} +\frac{32}{15}\zeta (6) \tau_2^{-5/2}+\ldots.\ee
If $\tilde\mu \neq 0$, we can still constrain the genus 3 coefficient $a^{(8)}_3$ using \C{meqn}. Given the genus 0 amplitude in \C{tree}, expanding
\be \label{tildeq}\tilde{q} = 2\zeta (7) \tau_2^{7/2} +a^{(8)}_3 \tau_2^{-5/2} +\ldots,\ee
convoluting \C{tildeq} with $E_{7/2}$ and setting $s=7/2$ in \C{meqn}, we get that
\be \zeta (7) \Big( \frac{32}{5} \zeta (6) - 3a^{(8)}_3\Big) =\ldots,\ee 
leading to $a^{(8)}_3 \sim \zeta (2)^3$ for the part we have calculated. Thus we find that
\be \label{a83} a^{(8)}_3 \sim \zeta (2)^3. \ee

From \C{f10}, for generic $\omega_i$, we get that
\be \label{f102}
f^{(10)}_i \sim \xi_i \zeta (5) \Big(\zeta (3) \tau_2^2 +\zeta (2) \Big) \tau_2^2 +\xi_i \zeta (4) \Big( \zeta (3) + \zeta (2) \tau_2^{-2}\Big) +\psi_i \zeta (4) \Big( ({\rm ln } \tau_2)^2 +{\rm ln} \tau_2 + 1\Big).\ee
The first term involves analytic terms which are genus 0 and genus 1 contributions, which are consistent with \C{tree} and~\cite{Green:2008uj}, while the non--analytic terms are  genus 2 contributions which is consistent with unitarity~\cite{Green:2006gt}. The second term yields analytic genus 2 and 3 contributions $\sim \zeta (3)\zeta (4)$ and $\sim \zeta (2) \zeta (4) \tau_2^{-2}$ respectively, in agreement with~\cite{Green:2008bf}. 

What if $\omega_i =6$ in \C{f10}? This yields genus 3 contributions not coming from the source terms. Then the equation
\be 4 \tau_2^2 \frac{\p^2}{\p \tau \p \bar\tau} r = 6 r + \xi E_{3/2} E_{5/2} + \psi Y^2\ee
is solved by
\be r(\tau,\bar\tau) = \frac{c_3}{\tau_2^2} -\frac{32}{25} \xi \zeta (2) \zeta (4) \tau_2^{-2}{\rm ln} \tau_2 +\ldots,\ee
and thus $\xi =0$ by unitarity. In this case, $c_3$ is not constrained by, for example, \C{meqn} and one needs to know non--perturbative information about $Y$.

The genus 5 contribution to $f^{(10)}_i \sim \zeta (10) \tau_2^{-6}$, as we are considering the $D^{10} t_8 t_8 R^4$ interaction (see \C{g5}).

\subsection{The $D^{14} \mathcal{R}^4$ and $D^{16} \mathcal{R}^4$ interactions}

The $D^{14} \mathcal{R}^4$ and $D^{16} \mathcal{R}^4$ couplings appear as source terms in the Poisson equation for the $D^{24} \mathcal{R}^4$ coupling.
The $D^{14} \mathcal{R}^4$ interaction has the unique spacetime structure given by $(s^2 + t^2 + u^2)^2 (s^3 + t^3 + u^3) \mathcal{R}^4$, and hence its coupling $f^{(14)}$ can be written as
\be f^{(14)} = \sum_i f^{(14)}_i.\ee 
However, the $D^{16} \mathcal{R}^4$ interaction has two distinct spacetime structures given by $(s^2 + t^2 + u^2)(s^3 + t^3 + u^3)^2 \mathcal{R}^4$ and $(s^2 + t^2 + u^2)^4\mathcal{R}^4$, and thus there are two couplings $f^{(16)}_i$ ($i=1,2$) given by
\be f^{(16)}_i = \sum_j f^{(16)}_{i,j}.\ee

Now $f^{(14)}_i$ satisfies the Poisson equation
\be \label{f14}4\tau_2^2 \frac{\p^2}{\p \tau \p \bar\tau} f^{(14)}_i = \underline\lambda_i f^{(14)}_i + E_{3/2} \Big( \sum_j \underline\alpha_{ij} f^{(8)}_j + \underline\beta_i E_{3/2}Y \Big) + \underline\gamma_i Y f^{(6)}  + \underline\theta_i E_{5/2}^2,\ee
while $f^{(16)}_{1,i}$, for example, satisfies 
\be \label{f16}4\tau_2^2 \frac{\p^2}{\p \tau \p \bar\tau} f^{(16)}_{1,i} = \underline\s_i f^{(16)}_{1,i} + E_{3/2} \Big( \sum_j \underline\omega_{ij} f^{(10)}_j + \underline\xi_i E_{3/2} E_{5/2} + \underline\psi_i Y^2 \Big) + \sum_j \underline\kappa_{ij} Y f^{(8)}_j  + \underline\Upsilon_i E_{5/2} f^{(6)}.\ee

Based on known results of perturbative string amplitudes and given the source terms $E_{3/2} f^{(8)}$,  $E_{3/2} f^{(10)}$ and $Y f^{(8)}$ in \C{f14} and \C{f16}, there can be extra source terms in these equations. First let us consider such terms which can contribute to \C{f14}. The $E_{3/2} f^{(8)}$ source term involves $f^{(8)}$ in $S^{(7)}$, which also includes the $D^6 \mathcal{R}^5, D^4 \mathcal{R}^6, \ldots$ interactions. The couplings of these interactions, multiplied by $E_{3/2}$, can also possibly contribute to the source terms in  \C{f14}. Note that these non--BPS couplings need not be the same as $f^{(8)}$.    
Similarly in \C{f16}, we can have other source terms because $f^{(10)}$, the $D^{10} \mathcal{R}^4$ coupling, is at the same order in the derivative expansion as the $D^8 \mathcal{R}^5$ coupling, for example.

In fact a part of the $D^8  \mathcal{R}^5$ interaction is in the same supermultiplet as the $D^{10} \mathcal{R}^4$ interaction. However a part of the $D^8 \mathcal{R}^5$ interaction with a specific spacetime structure is not, and we call its coupling $f^{(8;5)}$. Thus there can be a source term of the form $E_{3/2} f^{(8;5)}$ in \C{f16}. Now $f^{(8;5)}$ is a modular form which has vanishing genus zero amplitude, but a non--vanishing genus one amplitude~\cite{Green:2013bza}. A part of the genus one amplitude $\sim \zeta (2) \zeta (5)\tau_2^2$ just like the $D^{10} \mathcal{R}^4$ interaction given the modular invariant integrand. However, there are extra contributions to the genus one amplitude given by
\be \label{extra} \Upsilon = \frac{24\pi}{5}\int_{\mathcal{F}} \frac{d^2 \Omega}{\Omega_2^2} \Big(-\frac{1}{4}D_{222}^\p - D_{1111}^{''\p} \Big)\ee 
in the notation of~\cite{Green:2013bza}, coming from expressions involving the derivatives of the worldsheet scalar propagators, which have a different modular invariant integrand compared to the other contributions. It would be interesting to see what contribution it  gives to $f^{(8;5)}$.

\section{Some partial contributions to higher genus $D^8 \mathcal{R}^4$ and $D^{12} \mathcal{R}^4$ amplitudes}
 
In determining the analytic part of the genus five $D^{24} \mathcal{R}^4$ interaction upto  transcendentality 19, we need the coefficients of the analytic parts of the genus 5 contributions to the $D^8 \mathcal{R}^4$ and $D^{12} \mathcal{R}^4$ amplitudes in \C{genus5}, which are denoted $a^{(8)}_5$ and $a^{(12)}_5$ respectively. We shall now show, using \C{meqn} that $ a^{(8)}_5 \sim \zeta(2)^5$ and $a^{(12)}_5 \sim \zeta (2)^5 +\zeta (2)^5 \zeta (3)$ upto the order in transcendentality we need. In determining $a^{(12)}_5$ to the required order, we also need the expression for $a^{(8)}_6$, the coefficient of the analytic part of the genus 6 contribution to the $D^{8} \mathcal{R}^4$ interaction. This is also needed in section 4.7 in calculating the genus 6 contribution to the $D^{30} \mathcal{R}^4$ interaction.  In the analysis below, we shall drop all numerical factors as they are irrelevant for our analysis.

\subsection{The genus five and six $D^8 \mathcal{R}^4$ amplitude}

In order to calculate $a^{(8)}_5$, we consider the source terms in \C{18more} of $O(\tau_2^{-8})$. Neglecting various coefficients, they are given by
\be \Big( \zeta (2)^7 + a^{(8)}_5 \zeta (2)^2 + a^{(10)}_6 \zeta (2) {\rm ln} \tau_2\Big) \tau_2^{-8},\ee
where $a^{(10)}_6$ is the analytic part of the genus six $D^{10} \mathcal{R}^4$ coupling. Thus, for generic $\hat\lambda_i$ in \C{18more}, we have that
\be \label{eqnone}f^{(18)}_{1,i} \sim \Big( \zeta (2)^7 + a^{(8)}_5 \zeta (2)^2 + a^{(10)}_6 \zeta (2) \Big) \tau_2^{-8}+\ldots. \ee
Now convoluting \C{eqnone} with $E_9 = 2\zeta (18) \tau_2^9 + 2\sqrt{\pi}
\Gamma(17/2)\Gamma(9)^{-1}\zeta(17) \tau_2^{-8}+\ldots$, and using \C{meqn}, we get that
\be \zeta (2)^9 \Big( a^{(8)}_5 \zeta (2)^2 +a^{(10)}_6 \zeta (2) +\zeta (2)^7\Big) \sim \ldots,\ee
leading to 
\be \label{g58} a^{(8)}_5 \sim \zeta (2)^5, \quad a^{(10)}_6 \sim \zeta (2)^6.\ee

In calculating $a^{(8)}_6$, we consider \C{f14} which is the $D^{2k} \mathcal{R}^4$ interaction with the least number of derivatives which has $f^{(8)}_i$ as a source term. Taking $\underline\lambda_i = 90$, we see that \C{f14} has source terms at $O(\tau_2^{-9})$ given by
\be \label{eqnstep} \Big( \zeta (2)^7 + \zeta (2) a^{(8)}_6 +\zeta (3) a^{(8)}_7 \Big)\tau_2^{-9}\ee 
where $a^{(8)}_7$ is the analytic part of the genus seven $D^{8} \mathcal{R}^4$ coupling.
The $\zeta (2)^7 \tau_2^{-9}$ term in \C{eqnstep} comes from the genus seven $D^{14} \mathcal{R}^4$ amplitude using \C{genusn}.
Convoluting \C{eqnstep} with $E_{10} = 2\zeta (20) \tau_2^{10} + 2\sqrt{\pi}
\Gamma(19/2)\Gamma(10)^{-1}\zeta(19) \tau_2^{-9}+\ldots$, and using \C{meqn}, we get that
\be \zeta (2)^{10} \Big( \zeta (2)^7 + \zeta (2) a^{(8)}_6  + \zeta (3) a^{(8)}_7 \Big) \sim \ldots,\ee
leading to\footnote{At leading order in the transcendentality, $a^{(8)}_7 \sim \zeta (2)^7$ which gives an additional contribution $\sim \zeta (2)^6 \zeta (3)$ to $a^{(8)}_6$. However, we an ignore this term because it gives contributions which have more transcendentality compared to the terms we are keeping. }
\be \label{a86} a^{(8)}_6 \sim \zeta (2)^6. \ee

\subsection{The genus five $D^{12} \mathcal{R}^4$ amplitude}

The genus five $D^{12} \mathcal{R}^4$ amplitude is given by $a^{(12)}_5 \tau_2^{-11/2}$, which has not been determined in \C{ans6loop}. In order to calculate $a^{(12)}_5$, we consider the source terms in \C{18more} of $O(\tau_2^{-6})$, which are are given by
\be \Big( \zeta (2)^6 +\zeta (2)^6 \zeta (3) + a^{(12)}_5 \zeta (2) + a^{(8)}_6 \zeta (5) \Big) \tau_2^{-6},\ee
where $a^{(8)}_6$ is the analytic part of the genus six $D^{8} \mathcal{R}^4$ coupling. As before, for generic $\hat\lambda_i$ in \C{18more}, we have that
\be \label{eqntwo}f^{(18)}_{1,i} \sim \Big(\zeta (2)^6 +\zeta (2)^6 \zeta (3) + a^{(12)}_5 \zeta (2) + a^{(8)}_6 \zeta (5)  \Big) \tau_2^{-6}+\ldots. \ee
Convoluting \C{eqntwo} with $E_7 = 2\zeta (14) \tau_2^7 + 2\sqrt{\pi}
\Gamma(13/2)\Gamma(7)^{-1}\zeta(12) \tau_2^{-6}+\ldots$, and using \C{meqn}, we get that
\be \zeta (2)^7 \Big( \zeta (2)^6 +\zeta (2)^6 \zeta (3) + a^{(12)}_5 \zeta (2) + a^{(8)}_6 \zeta (5)\Big) \sim \ldots,\ee
leading to 
\be \label{g512} a^{(12)}_5 \sim \zeta (2)^5 + \zeta (2)^5 \zeta (3),\ee
 on using \C{a86} and keeping only terms upto transcendentality 13 in \C{g512}.

Now from \C{g58}, \C{a86} and \C{g512}, we get new contributions to the list in \C{hg}. They are given by

(i) Genus 5: \be \zeta (2)^5 \zeta (3) \tau_2^{-4}, \quad \zeta (2)^5 \zeta (5) \tau_2^{-4}, \quad \zeta (2)^5 \zeta (3) \zeta (5) \tau_2^{-4} \non \ee

(ii) Genus 6: \be\zeta (2)^6 \zeta (5)\tau_2^{-6} \non \ee

(iii) Genus 8: \be \label{newadd}\zeta (2)^8 \tau_2^{-10}. \ee

Hence we see that the constraints are stringent enough to produce various multi--loop string amplitudes.

So far as analyzing the structure of higher genus amplitudes is concerned, one should be able to see this structure and much beyond, coming out directly from the four graviton amplitude in maximal supergravity beyond 2 loops, generalizing the results of~\cite{Green:2008bf} and using~\cite{Bern:2007hh,Bern:2009kd}.



\providecommand{\href}[2]{#2}\begingroup\raggedright\endgroup


\end{document}